\documentclass[usenatbib]{mn2e}

\usepackage{epsfig}
\usepackage{mathrsfs}

\def\Lya{Ly$\alpha$~} 
\def\Lyb{Ly$\beta$~}

\def\LCDM{$\Lambda$CDM~}
\def\HI{\hbox{H~$\rm \scriptstyle I\ $}}
\def\HII{\hbox{H~$\rm \scriptstyle II\ $}} 
\def\HeI{\hbox{He~$\rm \scriptstyle I\ $}}
\def\HeII{\hbox{He~$\rm \scriptstyle II\ $}}
\def\HeIII{\hbox{He~$\rm \scriptstyle III\ $}} 
\def\MgII{\hbox{Mg~$\rm \scriptstyle II\ $}} 

\title[Nature and evolution of highly ionized near-zones]{The nature
  and evolution of the highly ionized near-zones in the absorption
  spectra of $z \simeq 6$ quasars}

\author[J.S. Bolton \& M.G. Haehnelt] {James S.
  Bolton$^{1}$\thanks{E-mail:jsb@ast.cam.ac.uk} \& Martin G.
  Haehnelt$^{1,2}$ \thanks{E-mail:haehnelt@ast.cam.ac.uk}\\
  $^1$ Institute of Astronomy, University of Cambridge, Madingley Road, Cambridge,  CB3 0HA \\
  $^2$Kavli Institute for Theoretical Physics, Kohn Hall, UCSB, Santa Barbara, CA 93106, USA\\}

\begin{document}

\date{11 October 2006}

\maketitle

\label{firstpage}

\begin{abstract}
  We use state-of-the-art hydrodynamical simulations combined with a
  $1$D radiative transfer code to assess the extent to which the
  highly ionized regions observed close to $z \simeq 6$ quasars, which
  we refer to as near-zones, can constrain the ionization state of the
  surrounding IGM. We find the appearance in \Lya absorption of a
  quasar \HII ionization front expanding into a neutral IGM can be
  very similar to a classical proximity zone, produced by the
  enhancement in ionizing flux close to a quasar embedded in a highly
  ionized IGM. The observed sizes of these highly ionized near-zones
  and their redshift evolution can be reproduced for a wide range of
  IGM neutral hydrogen fractions for plausible values of the
  luminosity and lifetime of the quasars.  The observed near-zone
  sizes at the highest observed redshifts are equally consistent with
  a significantly neutral and a highly ionized surrounding IGM.
  Stronger constraints on the IGM neutral hydrogen fraction can be
  obtained by considering the relative size of the near-zones in the
  \Lya and \Lyb regions of a quasar spectrum.  A large sample of high
  quality quasar absorption spectra with accurate determinations of
  near-zone sizes and their redshift evolution in both the \Lya and
  \Lyb regions should confirm or exclude the possibility that the
  Universe is predominantly neutral at the highest observed redshifts.
  The width of the discrete absorption features in these near-zones
  will contain important additional information on the ionization
  state and the previous thermal history of the IGM at these
  redshifts.

\end{abstract}
 

\section{Introduction}

Unravelling the epoch of hydrogen reionization is one of the major
remaining goals of observational cosmology. The discovery of extended
high opacity regions in the absorption spectra of $z\ga 6$ quasars has
provoked an intense discussion as to whether or not the tail end of
the reionization epoch has been detected.  Following the three year
WMAP results, the complementary constraints on the start and duration
of the reionization epoch from cosmic microwave background data are
now actually very weak (\citealt{Page06,Spergel06}). An extended
period of reionization starting at $z\sim 15$ and a rather rapid
reionization around $z\sim 6$ both appear to be consistent with the
data.  Much hope for further progress is based on upcoming experiments
to detect high-redshift $21\rm~cm$ emission/absorption
(\citealt{Scott90,Madau97,Shaver99,Tozzi00}).  Currently, however, the
study of Lyman series absorption in the spectra of quasars
(\citealt{Fan02,Fan06t}) and gamma-ray bursts
(\citealt{CiardiLoeb00,BarkanaLoeb04,Totani05}) are the key
observational probe of the hydrogen reionization epoch.  The absence
of a Gunn-Peterson trough (\citealt{GunnPeterson65}) in quasar spectra
at $z<5.5$ shows unambiguously that the volume weighted neutral
hydrogen fraction in the intergalactic medium (IGM) is less than one
part in $10^{4}$ (\citealt{Fan02,Fan06t}).  Hydrogen reionization is
therefore definitely complete at this time.  However, at $z\simeq6$
the detection of dark absorption troughs in quasar spectra indicates
the IGM neutral hydrogen fraction is increasing rapidly with look back
time
(\citealt{Djorgovski01,Becker01,Songaila02,Pentericci02,White03,Fan06t}).
Whether a large fraction of the volume is still neutral at this time
is, however, rather controversial. \cite{Fan06t} obtain a lower limit
on the volume weighted neutral hydrogen fraction of $f_{\rm HI} \geq
10^{-3.5}$ at $z \geq 6$ from a detailed analysis of the Lyman series
absorption in quasar spectra.  There are also claims for an
acceleration in the upwards evolution of the \Lya optical depth at
$z\simeq 6$ (\citealt{Cen02,Fan02,White03,Fan06t}) which may occur
during the epoch when individual \HII regions overlap
(\citealt{Gnedin00,Gnedin04}).  Other authors, however, argue that the
data is consistent with a smooth evolution due to the gradual
thickening of the \Lya forest (\citealt{Songaila04,Becker06b}).  Large
variations in the transmitted flux along different lines-of-sight
further complicate the picture.  Large Lyman series troughs are
detected in several quasar spectra at $z\simeq 6$ (\citealt{Fan06t}).
Yet the spectrum of $\rm J1148+5251$ at $z=6.42$ exhibits patchy
transmission, indicating the IGM is highly ionized in at least some
locations (\citealt{White03,White05,OhFurlanetto05}).  This may be
indicative of spatial fluctuations in the UV background which are
expected at the end of reionization (\citealt{Wyithe06,Fan06t}), but
may also be attributed to the fluctuations of the IGM density alone
(\citealt{Lidz06,Liu06}).  It is thus currently unclear whether the
IGM is highly ionized, substantially neutral or indeed a combination
of both just above $z=6$, depending on the line-of-sight considered.
Answering this question conclusively will have important implications
for the nature of hydrogen reionization.

The main limitation of Lyman series absorption as a probe of the IGM
neutral hydrogen fraction is it ceases to be sensitive at $f_{\rm HI}
\ga 10^{-3}$.  Other approaches to probing the IGM neutral hydrogen
fraction at $z \geq 6$ include analysing metal absorption systems
(\citealt{Oh02,FurlanettoLoeb03,Becker06}), the luminosity function of
\Lya emitting galaxies (\citealt{MalhotraRhoads04,Hu05,Kashikawa06}),
and the dark gap width distribution
(\citealt{PaschosNorman05,Gallerani05}) although all have their
associated interpretative uncertainties (see \citealt{Fan06rv} for an
recent review of results from all these techniques).  

In this work we shall concentrate on yet another probe of the neutral
hydrogen fraction of the high-redshift IGM.  Recently, claims that the
IGM has a large neutral hydrogen fraction at $z>6$ have been made based
on the observed sizes of the small regions of transmitted flux observed
in quasar absorption spectra just blue-ward of the quasar's redshift
(\citealt{Wyithe04,MesingerHaiman04,Mesinger04,Wyithe05,Yu05a}).  In
the regions close to the quasar intergalactic hydrogen is expected to
be highly ionized.  The typical sizes of these regions observed in
\Lya absorption at $z\simeq6$ are $5-10$ proper Mpc
(\citealt{Fan06t}).  Throughout this paper we shall refer to these
regions as near-zones.  This accounts for the possibility that these
regions may be either the observational signature of quasar \HII ionization
fronts expanding into a significantly neutral IGM, or classical
proximity zones due to enhanced ionizing flux near the quasar
(\citealt{Bajtlik88}).

\cite{Wyithe05} use the size of spectroscopically identified
near-zones observed around seven quasars at $6 \leq z \leq 6.42$ to
infer $f_{\rm HI} \geq 0.1$.  An alternative analysis by
\cite{MesingerHaiman04} uses the difference between the sizes of the
near-zones identified in the \Lya and \Lyb regions of the spectrum of
a single quasar at $z=6.28$ to obtain $f_{\rm HI} \geq 0.2$.  They
attribute the slightly smaller size of the \Lya near-zone to the
presence of a damping wing from a significantly neutral IGM
(\citealt{MiraldaEscudeRees98,MiraldaEscude98}). If these results are
correct, they would imply a very rapid increase in the neutral
hydrogen fraction over a short redshift interval, from $f_{\rm HI}
\sim 10^{-4}$ at $z=5.5$ to $f_{\rm HI} \geq 10^{-1}$ at $z \sim 6.4$.
In contrast, by comparing the relative sizes of \Lya near-zones
observed around $16$ quasars, \cite{Fan06t} conclude the IGM neutral
hydrogen fraction increases by a factor of $\sim 14$ assuming the
near-zone sizes scale as $(1+z)^{-1}f_{\rm HI}^{-1/3}$ over the same
redshift range.  They find no strong evidence for a significantly
neutral IGM at $z=6.4$.

These results appear to be in direct contradiction with each other.
However, there are a number of uncertainties which may explain the
origin of these different conclusions.  Using the properties of
spectroscopically identified near-zones to determine the IGM neutral
hydrogen fraction requires assumptions about the ionizing luminosity
and lifetime of quasars, as well as the density distribution of the
surrounding IGM.   The clumpy nature of the IGM will produce large
  variations in near-zone sizes from one line-of-sight to the next,
  with the underdense regions in the IGM setting the observed sizes of
  these regions ({\it e.g.} \citealt{OhFurlanetto05}). Further uncertainties
arise from the possibility of fainter ionizing sources clustered
around the quasar and the exact details of the radiative transfer
through the inhomogeneous IGM.  The sizes of the spectroscopically
identified near-zones themselves are also uncertain.  They are
determined from generally noisy, moderate or low resolution spectra
and the systemic redshift of the quasar is usually not known to a high
degree of accuracy.  It is also not immediately clear whether the
observed sizes of near-zones correspond to the full extent of the
regions impacted by the ionizing radiation of the quasars ({\it i.e.}
the region behind the quasar \HII ionization front), a key assumption
when inferring the IGM neutral hydrogen fraction from the sizes of
near-zones.

In this paper we investigate these issues in some detail.  We model
highly ionized \Lya and \Lyb near-zones around high redshift quasars
using a large hydrodynamical simulation combined with an accurate
scheme for radiative transfer through an inhomogeneous IGM of
primordial composition.  We shall consider the information one may
realistically infer about the ionization state of the IGM from the
observed sizes of Lyman series near-zones.  In
section~\ref{sec:stromprox} we develop a simple analytical model for
the sizes of the spectroscopically observed near-zones.  A brief
overview of our radiative transfer simulations is presented in
section~\ref{sec:sims}, and in section~\ref{sec:realdens} we examine
the dependence of \Lya and \Lyb near-zones sizes on the IGM neutral
hydrogen fraction, quasar age and ionizing photon production rate.  We
consider the evolution in \Lya near-zone sizes observed by
\cite{Fan06t} in section~\ref{sec:evolution} and present our
conclusions in section~\ref{sec:concs}.  Details of our radiative
transfer implementation and treatment of unresolved self shielded
clumps are found in the appendices.

Throughout this paper we adopt the following set of cosmological
parameters $(\Omega_{\rm m},\Omega_{\Lambda},\Omega_{\rm
  b}h^{2},h,\sigma_{8},n) = (0.26,0.74,0.024,0.72,0.85,0.95)$,
consistent with the combined analysis of three year WMAP and \Lya
forest data (\citealt{Viel06,Seljak06}), and a helium fraction by mass
of $Y=0.24$ ({\it e.g.} \citealt{OliveSkillman04}).  Unless otherwise
stated all distances referred to in the text correspond to proper
distances.

\section{\HII ionization fronts and the proximity effect} \label{sec:stromprox}
\subsection{The extent of the regions impacted by quasar ionizing radiation}

We firstly consider the case of an \HII ionization front (I-front) expanding into
a significantly neutral IGM.  Consider the idealised case
of an observer positioned along the line of sight to a
monochromatic ionizing source, isotropically emitting $\dot N$ Lyman
limit photons per unit time into a homogeneous, pure hydrogen IGM with
a volume weighted neutral hydrogen fraction $f_{\rm HI}= n_{\rm
  HI}/n_{\rm H}$.  If this observer could measure the exact size of
the \HII region around the source at various times $t$ during the
source lifetime, the \HII region expansion rate would be described by
(see appendix A for details)

\begin{equation} \frac{dR_{\rm ion}}{dt} = \frac{ \dot N - \frac{4}{3} \pi R_{\rm ion}^{3} 
    \alpha_{\rm HII} n_{\rm H}^{2}}{4 \pi R_{\rm ion}^{2} f_{\rm HI} n_{\rm H}},
    \label{eq:HIIfront} \end{equation}

\noindent
where $R_{\rm ion}$ is the distance of the \HII I-front from the
source, $\alpha_{\rm HII}$ is the recombination coefficient for
ionized hydrogen and $n_{\rm H}$ is the proper hydrogen number density
at a fixed baryonic density normalised by the cosmic mean, $\Delta =
\rho_{\rm b}/\bar \rho_{\rm b}$,

\begin{equation} n_{\rm H} = 7.0 \times 10^{-5} \Delta \left( \frac{1+z}{7} \right)^{3} \rm~cm^{-3}. \end{equation}

\noindent
For small values of $f_{\rm HI}$ the ``I-front'' obviously produces
only a very small change in the ionized hydrogen fraction within a
region of radius $R_{\rm ion}$. 

Solving equation~(\ref{eq:HIIfront}) for a source with age $t_{\rm Q}$
yields,

\begin{equation} R_{\rm ion} = R_{\rm S} \left[1 - \exp \left (-\frac{t_{\rm Q}}{f_{\rm HI} t_{\rm rec}}\right )\right]^{1/3},  \label{eq:stromgren} \end{equation}

\noindent
where $R_{\rm S}$ is the Str\"omgren radius, defined as

\begin{equation} R_{\rm S} = \left( \frac{3 \dot N}{4 \pi \alpha_{\rm HII} n_{\rm H}^{2}} \right)^{1/3}, \label{eq:reclimit} \end{equation}

\noindent
and $t_{\rm rec} = (n_{\rm H} \alpha_{\rm HII})^{-1}$ is the
recombination timescale.  If $ t_{\rm Q} \ll f_{\rm HI}t_{\rm rec}$
the solution to equation~(\ref{eq:HIIfront}) may be rewritten as

\[ R_{\rm ion} = \frac{4.2}{(\Delta f_{\rm HI})^{1/3}} \left( \frac{\dot N}{2 \times 10^{57} \rm~s^{-1}}
\right)^{1/3} \left( \frac{t_{\rm Q}}{10^{7} \rm~yrs} \right)^{1/3}
\]
\begin{equation} \hspace{7mm} \times  \left( \frac{1+z}{7} \right)^{-1} \rm~Mpc,  \label{eq:RHII} \end{equation}

\noindent
where we have adopted some fiducial values for a quasar embedded in a
uniform IGM.

Equation~(\ref{eq:HIIfront}) has been used extensively to model the
sizes of \HII regions around high redshift quasars ({\it e.g.}
\citealt{Shapiro87,Donahue87,Madau99,MadauRees00,CenHaiman00,White03,Wyithe04,Yu05a,Yu05b,Shapiro05}).
While equation~(\ref{eq:HIIfront}) is a reasonable approximation for
the expansion rate of an \HII I-front embedded in a significantly
neutral IGM, it does not predict the residual neutral hydrogen
fraction behind the I-front.  However, the sizes of \Lya and \Lyb
near-zones observed in quasar spectra will depend on the value of the
neutral hydrogen fraction behind the \HII I-front.  The value of
$R_{\rm ion}$ can therefore be different to the size of the near-zone
spectroscopically identified in Lyman series absorption.  For small
values of $f_{\rm HI}$, $R_{\rm ion}$ can be significantly larger than
the observed near-zone size.

\subsection{A simple model for the observed sizes of near-zones in Lyman series absorption}

The Gunn-Peterson optical depth (\citealt{GunnPeterson65}) resulting
from the scattering of redshifted \Lya photons by a uniform
distribution of neutral hydrogen can be written as,

\begin{equation} \tau(z) = \frac{\sigma_{\alpha} c n_{\rm H} f_{\rm HI}}{
H(z)},  \label{eq:GP} \end{equation}

\noindent
where $\sigma_{\alpha} = 4.48 \times 10^{-18} \rm~cm^{2}$ is the
scattering cross-section for \Lya photons, $H(z)$ is the Hubble
parameter and $c$ is the speed of light.

Now consider the effect of a luminous quasar on the surrounding IGM
behind its \HII I-front.  The extent of the observed near-zone in the
\Lya region of the quasar absorption spectrum, $R_{\alpha}$, will be
set by the limiting \Lya optical depth, $\tau_{\rm lim}= - \ln F_{\rm
  lim}$, at which the observer can clearly distinguish between the
presence and absence of transmitted flux.  However, as discussed by
\cite{Fan06t}, consistently defining the boundary of a near-zone is
difficult, and is complicated by fluctuations in the IGM density
distribution, the UV background and the low signal-to-noise and
resolution of the quasar spectrum.  In addition, at lower redshifts
the IGM has non-zero transmission outside of the near-zone, making the
definition of $R_{\alpha}$ ambiguous.  \cite{Fan06t} define
$R_{\alpha}$ as the extent of the region where $F_{\rm lim}>0.1$ when
the quasar spectrum has been smoothed to a resolution of $20\rm~\AA$.
The smoothing of the somewhat noisy spectra is very reasonable and a
common definition is obviously essential for comparing observed
spectra consistently. However the smoothing of the spectra can result
in an underestimate of the estimated near-zone sizes if associated
narrow transmission peaks beyond the region of continuous transmission
are obliterated, or in an overestimate if transmission from the IGM
ionized solely by the UV background is included.  We shall therefore
adopt a definition of $R_{\alpha}$ more suited to a discussion of the
general scaling laws for near-zone sizes in the absence of such
practical difficulties, and will come back to a detailed comparison
using the definition of \cite{Fan06t} later.  Here we do not apply any
smoothing and define the edge of the near-zone simply as corresponding
to the last pixel at which the spectrum drops below the normalised
flux detection limit of $F_{\rm lim} = 0.1$, corresponding to an
optical depth detection limit $\tau_{\rm lim} = 2.3$.  Note that this
definition will break down when transmission from the IGM ionized
solely by the metagalactic UV background becomes common.

Rearranging equation~(\ref{eq:GP}) and setting $\tau = \tau_{\rm lim}$ yields
the neutral hydrogen fraction required to reproduce the optical depth
detection limit,

\begin{equation} f_{\rm HI}^{\rm lim} \simeq 5.4 \times 10^{-6} \Delta_{\rm lim} \left( \frac{\tau_{\rm lim}}{2.3} \right)
  \left( \frac{1+z_{\alpha}}{7} \right)^{-3/2}, \end{equation}

\noindent
where $\Delta_{\rm lim}$ is the corresponding normalised baryon
density as fraction of the cosmic mean and $z_{\alpha}$ is the
redshift of the \Lya near-zone boundary.  Assuming the hydrogen behind
the quasar \HII I-front is highly ionized and in ionization
equilibrium, which is a good approximation in the vicinity of a
luminous quasar, the ionization rate per hydrogen atom needed to
reproduce the neutral hydrogen fraction at the detection limit is,

\begin{equation} \Gamma_{\rm HI}^{\rm lim} \simeq \frac{ \chi n_{\rm H} 
    \alpha_{\rm HII}(T_{\alpha}) }{ f_{\rm HI}^{\rm lim}},
    \label{eq:gamma} \end{equation}

\noindent 
where $\chi = 1.158$ if hydrogen and helium are fully ionized by the
quasar and $\chi = 1$ if hydrogen only is ionized, and $\alpha_{\rm
HII}(T_{\alpha})$ is the case A recombination coefficient evaluated at
temperature $T_{\alpha}$, which is the IGM temperature at
$z_{\alpha}$.  Converting this to a specific intensity gives

\[ J_{\rm HI}^{\rm lim} \simeq 1.34 \times 10^{-21} \Delta_{\rm lim}^{2} \left( \frac{T_{\alpha}}{2\times 10^{4} \rm~K} \right)^{-0.7}\left( \frac{\tau_{\rm lim}}{2.3} \right)^{-1} \] 
\begin{equation} \hspace{7mm} \times  \left( \frac{\alpha_{\rm s} +3}{4.5} \right) \left( \frac{1+z_{\alpha}}{7} \right)^{9/2} \rm~erg~s^{-1}~cm^{-2}~sr^{-1}~Hz^{-1}, \label{eq:JH1} \end{equation}

\noindent
assuming the quasar has power law spectrum with spectral index
$\alpha_{\rm s}=1.5$, $\chi = 1.158$ and $\alpha_{\rm HII} = 2.51
\times 10^{-13} \rm~cm^{3}~s^{-1}$, evaluated at $T_{\alpha} =2\times
10^{4}\rm~K$ (\citealt{Abel97}).  The largest observable size of the
\Lya near-zone will then correspond to the distance $R_{\alpha}^{\rm
  max} $ from the source beyond which the source ionizing radiation
field drops below $J_{\rm HI}^{\rm lim}$,

\begin{equation} R_{\alpha}^{\rm max} = \frac{1}{4\pi} \left( \frac{ \dot N h_{\rm p} \alpha_{\rm s}}
    {J_{\rm HI}^{\rm lim}} \right)^{1/2}, \label{eq:proximity}
    \end{equation}

\noindent
where $h_{\rm p}$ is Planck's constant and $\dot N$ is the number of
ionizing photons emitted by the quasar per unit time.  Combining
equations~(\ref{eq:JH1}) and (\ref{eq:proximity}) finally yields,

\[ R_{\alpha}^{\rm max} \simeq \frac{3.14}{\Delta_{\rm lim}} \left( \frac{\dot N}{2 
    \times10^{57} \rm~s^{-1}} \right)^{1/2} \left( \frac{T_{\alpha}}{2
    \times 10^{4}~\rm K} \right)^{0.35}\]

\begin{equation} \hspace{3mm}  \times  \left(\frac{\tau_{\rm
    lim}}{2.3}\right)^{1/2}\left( \frac{\alpha_{\rm s}^{-1}[\alpha_{\rm s}+3]}{3} \right)^{-1/2} \left(  \frac{1+z_{\alpha}}{7} \right)^{-9/4} \rm Mpc. \label{eq:Ralpha} \end{equation}

\begin{figure*}
  \centering \begin{minipage}{180mm} \begin{center}

      \psfig{figure=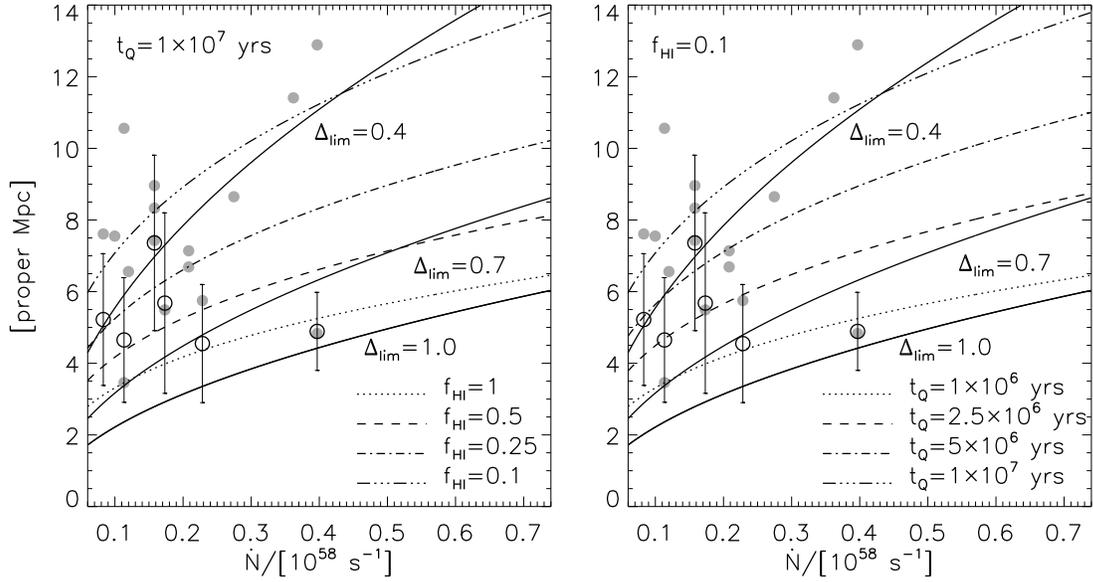,width=0.95\textwidth}
      \caption{ 
        Comparison of the observed sizes of \Lya near-zones as
      determined by  Fan et al. (2006b) (filled grey circles) and Wyithe et al.
        (2005) (open circles) to the two analytical models discussed
        in section~\ref{sec:stromprox}, which predict the extent of
        the \HII I-front and the size of the \Lya near-zone at $z=6$,
        respectively.  The sizes are plotted as a function of the
        number of ionizing photons emitted by the quasar per second,
        $\dot N$. The solid curves in both panels give
        $R_{\alpha}^{\rm max}$ for $\Delta_{\rm lim} = [0.4,0.7,1.0]$.
        {\it Left:} The series of dotted and dashed curves give
        $R_{\rm ion}$ assuming a uniform IGM with $\Delta=1$, a quasar
        age of $t_{\rm Q}=10^{7} \rm~yrs$ and an IGM neutral hydrogen
        fraction of $f_{\rm HI} = [1,0.5,0.25,0.1]$.  {\it Right:} The
        dotted and dashed curves give $R_{\rm ion}$ assuming a uniform
        IGM with $\Delta =1$, an IGM neutral hydrogen fraction of
        $f_{\rm HI} = 0.1$ and quasar ages of $t_{\rm Q} =
        [0.1,0.25,0.5,1] \times 10^{7} \rm~yrs$.}
      \label{fig:analytic}
      \end{center} \end{minipage}
\end{figure*}

\noindent
  Once the near-zone size reaches $R_{\alpha}^{\rm max}$, even if
  the \HII region around the quasar keeps growing, the residual
  neutral hydrogen in the ionized region beyond $R_{\alpha}^{\rm max}$
  will be sufficient to halt the advance of the observed near-zone
  edge.  Obviously the above definition for the maximum size of the
  near-zone obviously only makes sense if the amplitude of the
  metagalactic  UV background above the Lyman limit is smaller than 
  $J_{\rm HI}^{\rm lim}$.  Also note the size of the observed
  near-zone in the \Lya region will always be smaller than or equal 
  to $R_{\rm ion}$.  The important point to take from this simple
  argument is the different scaling of the two radii, $R_{\rm
  ion}\propto \dot N^{1/3}$ and $R_{\alpha}^{\rm max}\propto \dot
  N^{1/2}$, with the ionizing luminosity, and that $R_{\alpha}^{\rm
  max}$ is independent of the neutral hydrogen fraction of the IGM
  and the lifetime of the quasar. One may also derive a 
  similar expression for the maximum size of the observable \Lyb near-zone,

\begin{equation} R_{\beta}^{\rm max} \simeq  2.5  \left( \frac{T_{\beta}}{T_{\alpha}} \right)^{0.35}\left( \frac{1+z_{\beta}}{1+z_{\alpha}} \right)^{-9/4} R_{\alpha}^{\rm max}, \label{eq:Rbeta} \end{equation}

\noindent 
where $T_{\beta}$ is the IGM temperature at the edge of the \Lyb
near-zone at $z_{\beta}$.  

Equations~(\ref{eq:Ralpha}) and~(\ref{eq:Rbeta}) should be considered
as illustrative approximations only; radiative transfer through a
realistic IGM density distribution is required to address this fully.
A number of caveats should therefore be kept in mind.  The upper
limits for the near-zone sizes given by equations~(\ref{eq:Ralpha})
and~(\ref{eq:Rbeta}) do not take into account the possible effect of a
damping wing from a significantly neutral IGM, which reduces the size
of the \Lya near-zone relative to $\rm Ly\beta$
(\citealt{MesingerHaiman04}).  The presence of the dense,
self-shielded clumps responsible for observed Lyman-limit systems will
further reduce the upper limit for the observed sizes of the near-zone
compared to the estimates given by equations~(\ref{eq:Ralpha})
and~(\ref{eq:Rbeta}).  Additionally, the overlying lower redshift \Lya
forest should reduce the observed size of the near-zone in the \Lyb
region.  Note that for a predominantly neutral IGM, the sizes of the
observed \Lya and \Lyb near-zones will both trace the position of the
\HII I-front and should both be given by $R_{\rm ion}$,  whereas
if the IGM is highly ionized, the distinction between the 
near-zone and the general \Lya forest will become highly ambiguous.

However, assuming equation~(\ref{eq:Ralpha}) holds, for $R_{\rm
  ion}> R_{\alpha}^{\rm max}$ ($R_{\rm ion}> R_{\beta}^{\rm max}$)
the observed size of the \Lya (Ly$\beta$) near-zone becomes
independent of the neutral fraction of the surrounding IGM.  The
same is true if $R_{\rm ion}= R_{\rm S}$.  The timescale for the
\HII I-front to reach $R_{\alpha}^{\rm max}$ in a homogeneous IGM is
approximately,

\[ t \simeq 4.2 \times 10^{6} \Delta f_{\rm HI} \left ( \frac{R_{\alpha}^{\rm max}}{3.14 \rm~Mpc} \right)^{3} \left ( \frac{\dot N}{2 \times 10^{57} \rm~s^{-1}} \right)^{-1} \]
\begin{equation} \hspace{3mm} \times \left( \frac{1+z}{7} \right)^{3} \rm~yrs \label{eq:timescale} \end{equation}

\noindent
For a realistic clumpy IGM in the vicinity of a quasar this timescale
should be longer by a factor of a few, but for $f_{\rm HI} < 0.1$ this
is still considerably shorter than the canonical quasar lifetime of
$\sim 10^{7} \rm~yrs$ (\citealt{Martini04,Hopkins05}).  The time scale
for the \HII I-front to reach $R_{\beta}^{\rm max}$ is $\sim 16$ times
longer again.  The size of the \Lyb near-zone is therefore potentially
much more sensitive to the IGM neutral hydrogen fraction.  A similar
argument applies for the scaling of the observed near-zone sizes with
quasar age.

\subsection{Comparison of the analytical model to observational data}

The solid curves in both panels of figure~\ref{fig:analytic} show
$R_{\alpha}^{\rm max}$ and its dependence on the number of ionizing
photons emitted per second by the quasar, $\dot N$, using our fiducial
assumptions of $[\tau_{\rm lim},\alpha_{\rm s},T_{\alpha}, z_{\alpha}]
= [2.3,1.5,2\times 10^{4} \rm~K,6]$ in equation~(\ref{eq:Ralpha}).
The three curves shown are for $\Delta_{\rm lim}=[0.4,0.7,1]$.  In the
left panel the series of dotted and dashed curves show $R_{\rm ion}$
computed using equation~(\ref{eq:stromgren}) for $f_{\rm HI}
=[1,0.5,0.25,0.1]$ and a quasar lifetime of $t_{\rm Q}=10^{7}
\rm~yrs$.  In the right panel $R_{\rm ion}$ is shown for $t_{\rm
  Q}=[0.1,0.25,0.5,1] \times 10^{7}\rm~yrs$ and a fixed neutral
fraction of $f_{\rm HI} =0.1$.  When computing $R_{\rm ion}$, $\Delta
= 1$ has been assumed, although in reality $\Delta$ should be slightly
larger on average due to the clumpy nature of the IGM.  However, we
find this does not make any significant difference to our conclusions,
as we shall discuss later.

The filled grey circles in figure~\ref{fig:analytic} represent the
observed sizes of \Lya near-zones around quasars at $5.81 \leq z \leq
6.42$ as determined by \cite{Fan06t}.  In the instances where no \MgII
redshift was available, \cite{Fan06t} have applied a correction to the
quasar redshifts of $\Delta z = +0.02$ in order to take into account
the systematic redshift offset of high ionization lines. Note further
that, as discussed above, \cite{Fan06t} have smoothed the \Lya spectra
to a resolution of $20\rm~\AA$ before determining the near-zone sizes.
The open circles with error bars are \Lya near-zone sizes as
determined by \cite{Wyithe05} around a sub-sample of the same quasars
at $6 \leq z \leq 6.42$, adapted from their table 1.  \cite{Wyithe05}
used the \Lya near-zone sizes as a proxy for the extent of the quasar
\HII I-front, and determine the extent of the near-zones to be at the
redshifts where the transmission extending blueward of the \Lya
emission line becomes comparable to the noise.  The size of \Lya
near-zone around $\rm J1030+0524$ they quote was increased to
correspond to the size of the slightly larger \Lyb near-zone observed
around this quasar.  We have instead adopted $R_{\alpha} = 4.55 \pm
1.65$, with an error bar based on an uncertainty of $\Delta z = 0.03$
in the redshift of $\rm J1030+0524$ (\citealt{White03}).  We have also
omitted $\rm J1048+4637$ from the \cite{Wyithe05} sample, as this is a
BAL quasar which makes a meaningful measurement of the \Lya near-zone
size difficult (X.Fan, private communication).  For all except one of
the quasars ($\rm J1630+4012$, the lowest luminosity quasar) the
\cite{Wyithe05} values are consistent with the corresponding
determinations by \cite{Fan06t} within the error bars shown.  As
discussed by \cite{Fan06t} and later within this paper, the observed
near-zone sizes evolve rather rapidly with redshift.  The larger
near-zone sizes within the \cite{Fan06t} sample which are omitted in
the \cite{Wyithe05} data are for quasars at $z<6$.  For both data
sets, the conversion from ultraviolet AB absolute magnitudes,
$M_{1450}$, to $\dot N$ has been made using a generic broken power law
spectral energy distribution,

\begin{equation}
L({\nu})\propto \cases{\nu^{-0.5} &($1050<\lambda<1450\,$\AA),\cr
  \noalign{\vskip3pt}\nu^{-1.5} & ($\lambda<1050\,$\AA).\cr} \label{eq:SED}
\end{equation}

\noindent
All photons above $1\rm~Ryd$ are considered to be ionizing photons.
We have not plotted error bars for $\dot N$, although the values will
be somewhat uncertain depending on the exact spectrum of the quasars
({\it e.g.} \citealt{Yu05a}). 

As already discussed, the observed size of the near-zone in the \Lya
region should be the smaller of $R_{\alpha}^{\rm max}$ or $R_{\rm
  ion}$.  Clearly from figure~\ref{fig:analytic}, our simple
analytical argument suggests that the sizes of the observed \Lya
near-zones are consistent with both the expected extent of classical
proximity zones {\it and} an \HII I-front expanding into an IGM with a
significant neutral hydrogen fraction, for reasonable assumptions for
the quasar ionizing luminosity and lifetime.  The fact that
equation~(\ref{eq:stromgren}) and~(\ref{eq:Ralpha}) give similar
values for $R_{\rm ion}$ and $R_{\alpha}^{\rm max}$ is a coincidence
which explains the discrepant interpretation of the observed sizes of
near-zones in the literature.  Taking a combination of $0.1 \leq
f_{\rm HI} \leq 1 $ and $10^{6} \leq t_{\rm Q} \leq 10^{7} \rm~yrs$
results in an \HII region similar in size to the observed \Lya
near-zones, whereas equation~(\ref{eq:Ralpha}) reproduces the observed
sizes even for small $f_{\rm HI}$ or large $t_{\rm Q}$.  For example,
one can reasonably adopt values of $f_{\rm HI} = 10^{-3}$ and $t=5
\times 10^{7} \rm~yrs$, within the observationally determined limits
of $f_{\rm HI} \geq 10^{-3.5}$ (\citealt{Fan06t}) and $10^{6} \leq
t_{\rm Q} \leq 10^{8} \rm~yrs$ (\citealt{Martini04}).  For a quasar
luminosity of $\dot N=2\times 10^{57} \rm~s^{-1}$, assuming
$\Delta_{\rm lim}=0.4$ and $z_{\alpha}=6$, according to
equation~(\ref{eq:Ralpha}) one expects to see a \Lya near-zone of
$R_{\alpha} \simeq 7.9 ~\rm Mpc$, consistent with observed sizes.
However, equation~(\ref{eq:RHII}) predicts a value of $R_{\rm ion}
\simeq 22 \rm~Mpc$, much larger than any observed \Lya near-zone,
although in practise this will be slightly smaller due to the clumpy
nature of the IGM.  If the size of the observed near-zone is
nevertheless identified with $R_{\rm ion}$, the inferred IGM neutral
hydrogen fraction is erroneously large.

Therefore, the size of individual  near-zones in the \Lya region alone appears to
offer little in terms of constraining the ionization state of the
hydrogen in the surrounding IGM.  This will be confirmed by our
detailed numerical simulations including radiative transfer in
section~\ref{sec:realdens}. However, as we will show, this situation
improves if both the \Lya and \Lyb near-zones are considered in
tandem.

\section{Creating synthetic absorption spectra including near-zones} \label{sec:sims}
\subsection{Radiative transfer implementation}

Our $1$D radiative transfer implementation is based on an updated
version of the multi-frequency photon conserving algorithm used by
\cite{Bolton04}.  The advantage of this single ray-tracing approach is
that we may run many radiative transfer simulations with only modest
computational resources.  In addition, we achieve the high spatial and
temporal resolution which is required to accurately determine
post-reionization gas temperatures (\citealt{Bolton04,Tittley06}).
Technical details and the relevant resolution tests can be found in
appendices B and C, to which we refer the interested reader.

\subsection{Hydrodynamical simulations}

We combine our radiative transfer implementation with cosmological
density distributions drawn from a large hydrodynamical simulation.
The hydrodynamical simulation was run using the parallel TreeSPH code
{\tt GADGET-2} (\citealt{Springel05}).  The
simulation volume is a periodic box $60h^{-1}$ comoving Mpc in length
and contains $2 \times 400^{3}$ gas and dark matter particles.  Star
formation is included using the multi-phase model of
\cite{SpringelHernquist03} with winds disabled.  The simulation was
started at $z=99$, with initial conditions generated using the
transfer function of \cite{EisensteinHu99} and cosmological parameters
consistent with the combined analysis of three year WMAP and \Lya forest
data (\citealt{Viel06,Seljak06}), $(\Omega_{\rm
  m},\Omega_{\Lambda},\Omega_{\rm b}h^{2},h,\sigma_{8},n) =
(0.26,0.74,0.024,0.72,0.85,0.95)$.  The initial gas temperature was
set to $T=272.8 \rm~K$ and the gravitational softening length is
$3h^{-1}$ comoving kpc.  To check the numerical convergence of our
results we also run two further hydrodynamical simulations with lower
mass resolution.  The different resolution parameters for all three
simulations are listed in table~\ref{tab:sims}; all other aspects of
the lower resolution simulations are identical to the $400^{3}$ run.

\begin{table} 
  \centering 
  \caption{
    Hydrodynamical simulation resolution
    properties.  Listed in each column are the total gas particle
    number, the mass resolution per gas particle and the gravitational
    softening length. }

  \begin{tabular}{c|c|c} 
    \hline 
    Gas particle  & Gas particle         & Softening length \\   
    number        & mass [$M_{\odot}/h$] &  [comoving kpc$/h$] \\
    \hline
    $120^{3}$     & $1.61 \times 10^{9}$ & $10$ \\ 
    $240^{3}$     & $2.01 \times 10^{8}$ & $5$  \\ 
    $400^{3}$     & $4.34 \times 10^{7}$ & $3$  \\
    \hline
    
    \label{tab:sims}
  \end{tabular}
\end{table}

\begin{table} 
  \centering 
  \caption{
    Coordinates and masses of the five most massive haloes in the $400^{3}$ simulation 
    volume at $z=6$.  The halo coordinates are given in units of comoving kpc/$h$. }

  \begin{tabular}{c|c|c|c} 
    \hline 
    $x_{\rm pos}$     & $y_{\rm pos}$    &  $z_{\rm pos}$    & Halo mass [$M_{\odot}$/h]\\    
    \hline
     51030.5 & 46853.1 & 18807.6 & $2.65 \times 10^{12}$\\ 
     55886.6 & 49077.1 & 12091.1 & $1.42 \times 10^{12}$\\ 
     52392.1 & 48011.6 & 21267.5 & $1.33 \times 10^{12}$\\
     52061.7 & 47983.3 & 22023.3 & $1.25 \times 10^{12}$\\
     49415.3 & 47008.2 & 18859.4 & $1.05 \times 10^{12}$\\
    \hline
    
    \label{tab:haloes}
  \end{tabular}
\end{table}

A friends-of-friends halo finding algorithm was used to identify the
$5$ most massive haloes within the simulation volume in three
different snapshots at $z=[6.25,6.0,5.75]$.  The total masses (dark
matter, gas and stars) and coordinates of the haloes in the $z=6$
snapshot are listed in table~\ref{tab:haloes}.  These halo masses may
be slightly smaller than, or similar to, the expected mass of bright
quasar host haloes at $z \simeq 6$.  The typical comoving space
density of quasars at $z \simeq 6$ is $\sim 10^{-9}\rm~Mpc^{-3}$
(\citealt{Fan01}), which coincides with the expected abundance of dark
matter haloes with mass $\simeq 10^{13} M_{\odot}$
(\citealt{MoWhite02}).  However, recent observations of molecular gas
surrounding $\rm J1148+5251$ indicate its host halo mass may be closer
to $\simeq 10^{12} M_{\odot}$ (\citealt{Walter04}). Theoretical
arguments also suggest that the quasar occupation fraction of massive
haloes may be significantly less than unity, perhaps favouring host
haloes with lower average masses (\citealt{Volonteri06}).  In any
case, we have checked that our results depend only weakly on the halo
mass by running simulations centred on haloes of smaller mass.
  
At each redshift, five lines-of-sight where extracted around these
haloes, giving $15$ different lines-of-sight in total.  To construct
the density distributions for the radiative transfer runs we splice
together separate lines-of-sight drawn randomly from the simulation
volume. The first $30h^{-1}$ comoving Mpc of a spliced line-of-sight
is constructed using the density distribution drawn around one of the
massive haloes, in which we assume the quasar resides.  The remaining
$120h^{-1}$ comoving Mpc of the sight-line is made from two further
sight-lines drawn randomly from the simulation box at the same
redshift.  This gives a total length of $150h^{-1}$ comoving Mpc,
which is large enough to easily accommodate the observed sizes of \Lya
and \Lyb near-zones at $z\simeq 6$.  Each line-of-sight consists of
$3000$ pixels with a size of $9.9$ kpc at $z=6$.  This provides a good
degree of convergence for our radiative transfer simulations (see
appendix C for details).  For reference, the redshift extent of each
line-of-sight at $z=[6.25,6,5.75]$ is $\Delta z = [0.50,0.47,0.45]$.
Since the density distributions are drawn from a snapshot at a single
redshift, the gas density is rescaled by a factor of $(1+z)^{3}$ along
each spliced line-of-sight.

\begin{table} 
  \centering \caption{
    Spatially uniform UV background models used for
    setting the neutral hydrogen and helium fractions of the 
    surrounding IGM.  The first two
    columns list the specific intensity and ionization rate at the
    Lyman limit in units of $10^{-21}
    \rm~erg~s^{-1}~cm^{-2}~sr^{-1}~Hz^{-1}$ and $10^{-12} \rm~s^{-1}$
    respectively.  The last column lists the resulting neutral hydrogen
    fraction for mean cosmic density at $z=6$ assuming ionization
    equilibrium at $T=2 \times 10^{4}\rm~K$.  A power law spectrum 
    with  $\alpha_{\rm b}=1.5$ has
    been adopted.}

  \begin{tabular}{c|c|c|c} 
    \hline 
    Model & $J_{-21}^{\rm b}$  & $\Gamma_{-12}^{\rm b}$ & $f_{\rm HI}$ \\   
    \hline 
    0  &  0                   & 0                    & 1                     \\ 
    1  & $2.0 \times 10^{-5}$ & $5.6 \times 10^{-5}$ & $3.6 \times 10^{-1}$ \\
    2  & $1.0 \times 10^{-4}$ & $2.8 \times 10^{-4}$ & $7.3 \times 10^{-2}$ \\
    3  & $3.0 \times 10^{-4}$ & $8.5 \times 10^{-4}$ & $2.4 \times 10^{-2}$ \\ 
    4  & $1.0 \times 10^{-3}$ & $2.8 \times 10^{-3}$ & $7.3 \times 10^{-3}$ \\ 
    5  & $3.0 \times 10^{-3}$ & $8.5 \times 10^{-3}$ & $2.4 \times 10^{-3}$ \\ 
    6  & $1.0 \times 10^{-2}$ & $2.8 \times 10^{-2}$ & $7.3 \times 10^{-4}$ \\ 
    7  & $3.0 \times 10^{-2}$ & $8.5 \times 10^{-2}$ & $2.4 \times 10^{-4}$ \\ 
    8  & $6.0 \times 10^{-2}$ & $1.7 \times 10^{-1}$ & $1.2 \times 10^{-5}$ \\
    9  & $1.0 \times 10^{-1}$ & $2.8 \times 10^{-1}$ & $7.3 \times 10^{-5}$ \\
    10 & $2.0 \times 10^{-1}$ & $5.6 \times 10^{-1}$ & $3.6 \times 10^{-5}$ \\
    
    \hline
\label{tab:preion}

\end{tabular}
\end{table}

\subsection{Initial conditions}

We compute the transfer of radiation through an inhomogeneous hydrogen
and helium IGM from a source which emits $\dot N$ photons per second
above the \HI ionization threshold frequency, where

\[ \dot N = \int_{\nu_{\rm HI}}^{\infty} L_{\rm HI} \left( \frac{\nu}{\nu_{\rm HI}}
\right)^{-\alpha_{\rm s}}d\nu. \]

\noindent
The source is assumed to have a power law spectrum with index
$\alpha_{\rm s}$, typical of quasars, normalised by $L_{\rm HI}$ at
the \HI ionization threshold frequency, $\nu_{\rm HI}$.  We adopt
$\alpha_{\rm s}=1.5$ for this study.  The source is placed $10h^{-1}$
comoving kpc from the edge of the line-of-sight; all hydrogen and
helium within this radius is assumed to be fully ionized.  The
distribution of normalised densities $\Delta=\rho_{\rm b}/\bar
\rho_{\rm b}$ is kept constant during the radiative transfer
simulations, while the gas density evolves as $(1+z)^{3}$ as the
ionization front progresses along the line-of-sight. This should be a
reasonable approximation over a typical quasar lifetime.

We vary the lifetime of the source, $t_{\rm Q}$, the number of
ionizing photons emitted per second by the source, $\dot N$, and the
neutral hydrogen (and helium) fraction of the IGM.  To set the neutral
fractions in the IGM for each run, we assume the gas is in ionization
equilibrium with a spatially uniform UV background.  The different UV
background models we use are listed in table~\ref{tab:preion}, along
with the resulting neutral hydrogen fraction at mean density at $z=6$.
Note, however, that this does not necessarily correspond to the
resulting volume weighted neutral fraction in a simulated sight-line,
as on average these will be moderately overdense and have varying gas
temperatures, although it gives a rough guide to the expected neutral
hydrogen fraction for a given UV background.  All UV background models
assume a spectral index of $\alpha_{\rm b}=1.5$, although the
background spectrum may well be softer if it is dominated by galaxies.
This assumption does not have a significant effect on the results of
this paper apart from small changes in the neutral helium fraction and
the temperature of the IGM.  There should also be significant spatial
fluctuations in the UV background at $z\simeq 6$
(\citealt{Wyithe06,Fan06t}) but these will only be important at the
edge of the quasar near-zone itself and only if the flux from the
quasar and the metagalactic ionizing background are similar there. In
the latter case spatial fluctuations of the metagalactic ionizing
background may affect the measured near-zone sizes. 

\subsection{Constructing synthetic absorption spectra} \label{sec:spectra}

We construct synthetic spectra from the output of the radiative
transfer simulations as follows.  Each line-of-sight is rebinned to have
$N=4096$ pixels of proper width $\delta R$, each of which has a
neutral hydrogen number density $n_{\rm HI}$, temperature $T$,
peculiar velocity $v_{\rm pec}$ and Hubble velocity $v_{\rm H}$
associated with it.  The \Lya optical depth in each pixel is computed
assuming a Voigt line profile, such that the optical depth in pixel
$i$ corresponding to Hubble velocity $v_{\rm H}(i)$ is

\begin{equation} \tau_{\alpha}(i) = \frac{c \sigma_{\alpha}\delta R }{\pi^{1/2}}  
  \sum_{j=1}^{N} \frac{n_{\rm HI}(j)}{b(j)} H(a,x). \label{eq:voigt}
  \end{equation}

\noindent
Here $b = (2k_{\rm B}T/m_{\rm H})^{1/2}$ is the Doppler parameter,
$\sigma_{\alpha} = 4.48 \times 10^{-18} \rm~cm^{2}$ is the \Lya
scattering cross-section and $H(a,x)$ is the Hjerting function
(\citealt{Hjerting38})

\begin{equation} H(a,x) = \frac{a}{\pi} \int^{\infty}_{-\infty} \frac{e^{-y^{2}}}{a^{2} + 
    (x-y)^{2}} ~dy, \end{equation}

\noindent
where $x = [v_{\rm H}(i) - u(j)]/b(j)$, $u(j) = v_{\rm H}(j) + v_{\rm
  pec}(j)$, $a = \Lambda_{\alpha} \lambda_{\alpha} /4\pi b(j)$,
$\Lambda_{\alpha} = 6.265 \times 10^{8} \rm~s^{-1}$ is the damping
constant and $\lambda_{\alpha}=1215.67 \rm~\AA$ is the wavelength of
the \Lya transition.  For \Lyb optical depths we adopt the same
procedure, using instead $\sigma_{\beta} = 7.18 \times 10^{-19}
\rm~cm^{2}$, $\Lambda_{\beta} = 1.897 \times 10^{8} \rm~s^{-1}$ and
$\lambda_{\beta} = 1025.72 \rm~\AA$. In addition, for the \Lyb spectra
we also add the optical depth contribution from the \Lya forest at
lower redshift, $ z^{\prime} =
[\lambda_{\beta}(1+z)/\lambda_{\alpha}]-1$, where $z$ is the redshift
of the \Lyb pixels.  The total \Lyb optical depth is then
$\tau_{\beta}^{\rm tot}(z) = \tau_{\beta}(z) +
\tau_{\alpha}(z^{\prime})$.  In order to model the \Lya forest at
$z^{\prime}$, we use the standard technique of rescaling the optical
depth distribution of synthetic spectra, also drawn from our $400^{3}$
simulation at lower redshift, to reproduce the observed mean flux
$\langle F \rangle$ of the \Lya forest (see \citealt{Bolton05} for
details).  The values for the mean flux we adopt in this paper are
interpolated from \cite{Songaila04}: $\langle F \rangle =
[0.149,0.210,0.223]$ at $z^{\prime} = [5.12,4.91,4.70]$, corresponding
to $z=[6.25,6.0,5.75]$.

\begin{figure*}
  \centering
  \begin{minipage}{180mm}
    \begin{center}
   
     \psfig{figure=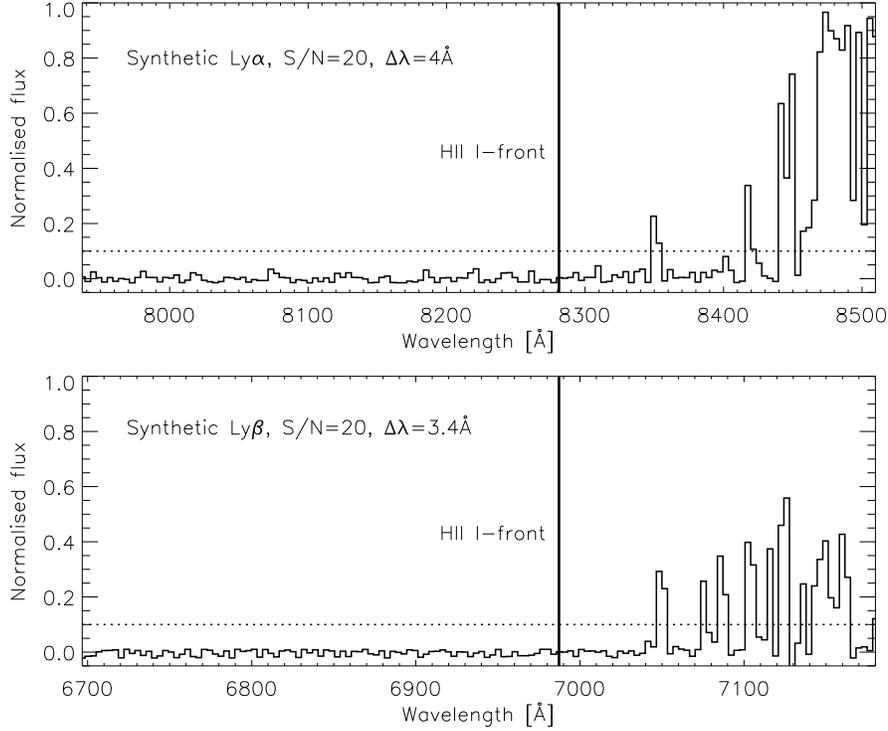,width=0.75\textwidth}
      \caption{
        Examples of the synthetic \Lya and \Lyb near-zone spectra used
        in our study.  The spectra are drawn from the line-of-sight
        data shown in figure~\ref{fig:los}.  The quasar is situated on
        the right of the diagram, and has an age $t_{\rm
          Q}=10^{7}\rm~yrs$ and ionizing luminosity $\dot N = 2 \times
        10^{57} \rm~s^{-1}$.  The surrounding IGM neutral hydrogen
        fraction is set using UV background model 4, which gives a
        volume weighted value of $f_{\rm HI} \sim 10^{-2}$. Here the
      \Lya spectrum has been rebinned to a resolution of $\sim
      4\rm~\AA$ per pixel and Gaussian distributed noise with a total
      signal to noise of $20$ per pixel has been added, typical of
      moderate resolution quasar spectra.  The \Lyb spectrum has been
      rebinned to have the same bin size as the \Lya spectrum in
      velocity space.  The solid line in each panel marks the extent
      of the quasar \HII region, some way ahead of the edge of
      spectroscopically identified \Lya and \Lyb near-zones.  These
      both lie at a distance of $\sim 8.3$ proper Mpc from the quasar.
      The dotted line shows the adopted flux threshold, $F_{\rm
        lim}=0.1$, used to identify the edge of the near-zone.}
      \label{fig:example} \end{center} \end{minipage}

\end{figure*}

\begin{figure*}
  \centering 
  \begin{minipage}{180mm}
    \begin{center}
      
      \psfig{figure=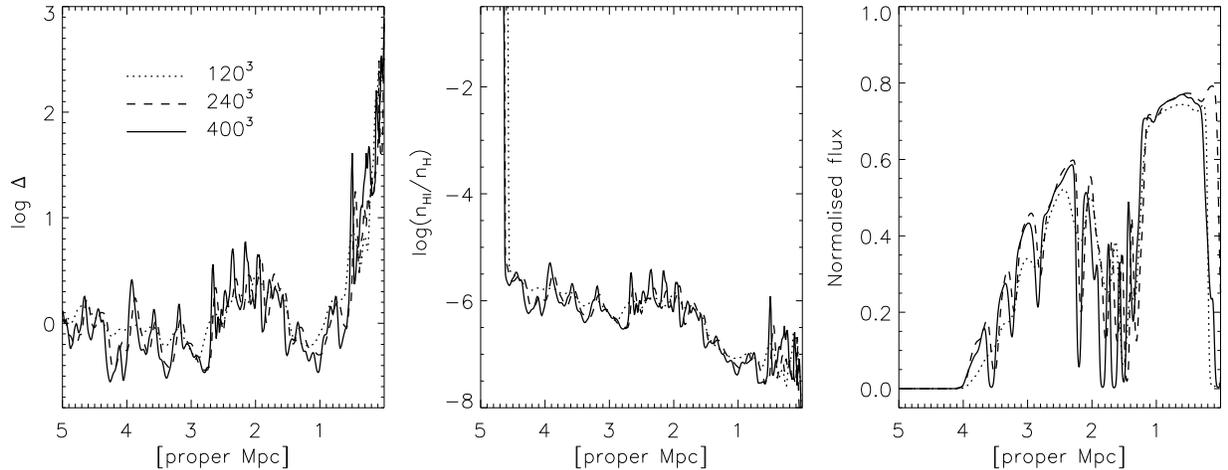,width=0.95\textwidth}
      \caption{  {\it Left:} Density distributions at $z=6$ drawn from
        hydrodynamical simulations with differing mass resolution.
        The dotted, dashed and solid lines correspond the $120^{3}$,
        $240^{3}$ and $400^{3}$ simulations respectively.  The
        corresponding mass resolutions are listed in
        table~\ref{tab:sims}. {\it Centre:} The neutral hydrogen
        number density normalised by the total hydrogen number density
        as predicted by the radiative transfer algorithm for a quasar
        within the massive halo situated on the right hand side of the
        density distribution.  {\it Right:} The resulting \Lya
        near-zone profile.}  \label{fig:hydrores} \end{center}
  \end{minipage}
\end{figure*}

For comparison to observational data, we must also mimic the typical
resolution and noise of quasar spectra at $z \simeq 6$.  The values
for these vary in the literature from one line-of-sight to the next.
We adopt a spectral resolution of $4\rm~\AA$ per pixel by rebinning
our simulated spectra and adding Gaussian distributed noise with a
total signal-to-noise ratio of $20$ per pixel at the continuum level
({\it e.g.} \citealt{Fan00}) and a constant read-out signal-to-noise
of $80$ per pixel.  Example synthetic \Lya and \Lyb spectra are shown in
figure~\ref{fig:example}.

\subsection{Hydrodynamical simulation resolution}

To simulate the propagation of ionizing radiation through an
inhomogeneous IGM, the structure of the IGM must be sufficiently
resolved to correctly model the number of ionizing photon absorption
events per unit volume ({\it e.g.}
\citealt{Kohler05,Iliev05,GnedinFan06}).  We test this using the
hydrodynamical simulations listed in table~\ref{tab:sims}.  We firstly
select one of our spliced lines-of-sight at $z=6$ from the $400^{3}$
simulation, and construct lines-of-sight from the same location in the
lower resolution simulations.  For all three lines-of-sight we then use
the radiative transfer code to model the propagation of ionizing
radiation from a quasar with $\dot N = 5 \times 10^{57} \rm~s^{-1}$
and $t_{\rm Q} = 10^{7} \rm~yrs$ into fully neutral hydrogen and
helium gas.  We subsequently extract the \Lya absorption
spectra.

The results of this test are shown in figure~\ref{fig:hydrores}.  From
left to right the panels show the density distribution, neutral
hydrogen fraction and the resulting \Lya near-zone profile drawn from
the radiative transfer runs in all three sight-lines.  The degree of
convergence in the \HII I-front position is good.  The
highest resolution simulation resolves a greater number of overdense
peaks, but the effect that this  has on the I-front position is
counteracted by better resolved voids.  The main difference is that
one sees more detailed structure in the higher resolution simulation
spectra.  Note also the smooth nature of the \Lya near-zone profile
due to the damping wing from the neutral IGM; the edge of the \Lya
near-zone lies behind the \HII I-front as a result ({\it e.g.}
\citealt{MesingerHaiman04}).  We conclude that the $400^{3}$
hydrodynamical simulation is sufficiently resolved for the purpose of
this work with one caveat; the very dense regions responsible for
Lyman limit absorption systems will not be sufficiently resolved.  We
will come back to this in section 5.


\section{Multi-frequency radiative transfer through cosmological density fields} \label{sec:realdens}
\subsection{An example line-of-sight}

\begin{figure*}
  \centering \begin{minipage}{180mm} \begin{center}

      \psfig{figure=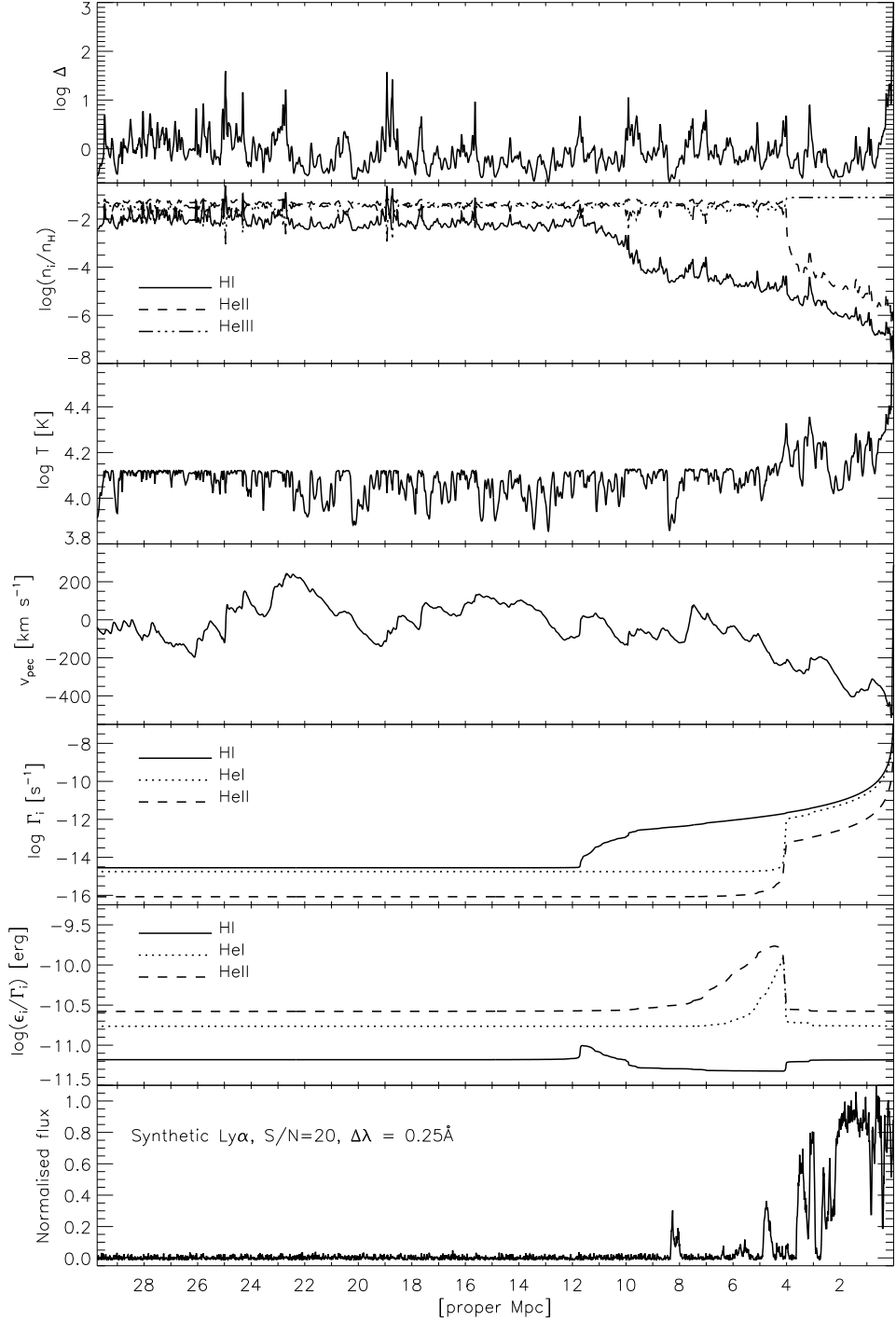,width=0.75\textwidth}
      \caption{ 
        An example line-of-sight used in our analysis of near-zone
        sizes.  The top panel shows the normalised density
        distribution along the line-of-sight drawn from a large
        hydrodynamical simulation at $z=6$ against distance from the
        quasar; the quasar host halo is on the right hand side of the
        diagram.  Subsequent panels show the output from the radiative
        transfer algorithm for a quasar of age $t_{\rm Q}=10^{7}
        \rm~yrs$ and ionizing luminosity $\dot N=2\times 10^{57}
        \rm~s^{-1}$. UV background model $4$ was used, producing an
        IGM with a volume weighted neutral hydrogen fraction $f_{\rm
          HI} \sim 10^{-2}$.  In descending order the panels display
        the H~$\rm \scriptstyle I$, \HeII and \HeIII number densities
        normalised by the total hydrogen number density, the gas
        temperature, the peculiar velocity (drawn from the
        hydrodynamical simulation), the H~$\rm \scriptstyle I$, \HeI
        and \HeII ionization rates and the energy input per ionization
        for H~$\rm \scriptstyle I$, \HeI and He~$\rm \scriptstyle II$.
        The last panel shows the resulting \Lya spectrum with a
        resolution and signal-to-noise resembling observed high
        quality, high resolution quasar spectra.}  \label{fig:los}
      \end{center} \end{minipage}
\end{figure*}

In figure~\ref{fig:example} we had shown synthetic absorption spectra
for an example line-of-sight at $z=6$.  Figure~\ref{fig:los} shows a
variety of physical properties along the same line-of-sight together
with a high resolution, high signal-to-noise version of the \Lya
region of the absorption spectrum.  The quasar is on the the right
hand side of the plot.  The scale on the horizontal axis corresponds
to the distance from the quasar in proper Mpc.  The top panel shows
the normalised density $\Delta$ along the line-of-sight.  In the
subsequent panels the output from the radiative transfer code is
shown.  In descending order the panels display the H~$\rm \scriptstyle
I$, \HeII and \HeIII number densities normalised by the total hydrogen
number density, the gas temperature, the peculiar velocity (drawn from
the hydrodynamical simulation), the H~$\rm \scriptstyle I$, \HeI and
\HeII ionization rates and the energy input per ionization for H~$\rm
\scriptstyle I$, \HeI and He~$\rm \scriptstyle II$.  The last panel
shows the resulting \Lya near-zone spectrum.  The ionizing source has
an age $t_{\rm Q}=10^{7} \rm~yrs$, luminosity $\dot N = 2\times
10^{57} \rm~s^{-1}$ and UV background model $4$ has been adopted.

There are a few interesting points to note about figure~\ref{fig:los}
before proceeding.  Firstly, the \HII I-front has reached $\sim
11.8\rm~Mpc$, which is somewhat smaller than the value of $R_{\rm
  ion}= 16.6 \rm~Mpc$ one predicts with equation~(\ref{eq:stromgren})
assuming $\Delta = 1$.  This difference is largely due to the clumpy
nature of the IGM and to a lesser extent the presence of helium.  
Adopting $\Delta = 2.2$, the actual mean normalised density along the
line-of-sight,   equation~(\ref{eq:stromgren})  predicts $R_{\rm ion}= 11.8
\rm~Mpc$, consistent with the simulation results. It is the mean
density of the gas which limits the   propagation of the I-front in 
this instance.  In comparison, the \HeIII I-front has reached the 
significantly smaller distance of $\sim 4.1$ Mpc from the source.  
The large difference in the extent of the
\HeIII and \HII I-fronts is due to the smaller fraction of \HI
compared to \HeII for the assumed metagalactic UV background.

The amount by which the quasar \HeIII I-front lags behind the \HII
I-front depends sensitively on the \HI to \HeII ratio in the
surrounding IGM.  Note that for a fully neutral IGM the extent of \HeIII and
\HII I-fronts should be very similar for typical quasar spectral index
(\citealt{Madau99}).  There is also evidence for a clear break in the
mean temperature of the IGM around $4 \rm~Mpc$, corresponding to the
\HeIII I-front position.  The higher temperature is due to radiative
transfer effects boosting the average energy per ionization when the
\HeII is reionized by the quasar (\citealt{AbelHaehnelt99,Bolton04}).
This effect becomes progressively larger for higher neutral helium
fractions.  Hence the widths of \Lya absorption features observed in
high resolution spectra of quasar near-zones ({\it e.g.}
\citealt{Becker05}) and their correlation (or lack of) with distance
from the quasar, combined with a lower limit to the extent of the \HII
I-front, contain interesting information on the ionization state
of hydrogen and helium in the IGM as well as the spectral index of the
source.  However, this may be complicated by the presence of high
temperature shocked gas near the quasar host halo.  
Detailed modelling  will be needed
before any firm conclusions can be drawn from observational data.

The second and third last panels in figure~\ref{fig:los} show the
behaviour of the combined quasar and metagalactic background radiation
fields at three frequencies, corresponding to the ionization
thresholds of H~$\rm \scriptstyle I$, \HeI and He~$\rm \scriptstyle
II$.  The quasar radiation field falls in intensity due to increasing
distance from the quasar and the decreasing photon mean free path,
until the ionization rate per atom/ion drops to the uniform value
adopted for the UV background model.  The filtering effect of
radiative transfer on the spectrum of the quasar radiation field is
clearly visible in the second last panel.  Higher energy photons have
a longer mean free path, boosting the average energy per ionization in
places where the gas is optically thick and the spectrum has
consequently hardened.  Note that the energy per ionization for \HI
actually decreases somewhat between the \HII and \HeIII I-fronts
because the photo-heating rate for \HI falls in this region as high
energy photons are soaked up by the optically thick helium.

\begin{figure*}
  \centering \begin{minipage}{180mm} \begin{center}
     
      \psfig{figure=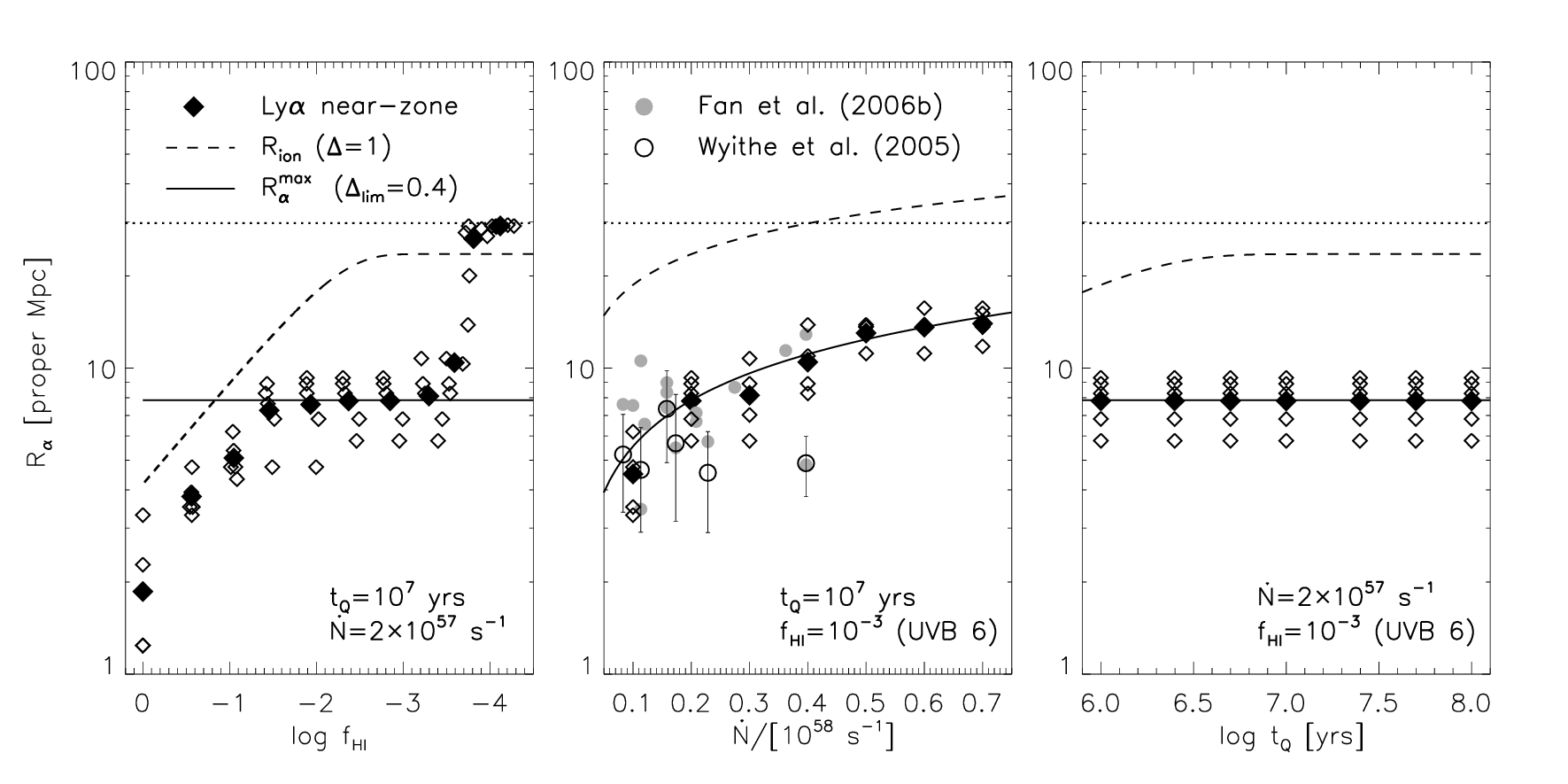,width=0.95\textwidth}
      \caption{ 
        The sizes of \Lya near-zones around quasars at $z=6$ computed
        from radiative transfer simulations in an inhomogeneous
        hydrogen and helium IGM.  In the left panel the IGM neutral
        hydrogen fraction is varied for a fixed quasar age and
        luminosity, in the central panel the quasar luminosity is
        varied for fixed quasar age and IGM neutral hydrogen fraction,
        and in the right panel the quasar age is varied for fixed
        quasar luminosity and IGM neutral hydrogen fraction.  The
        values of the fixed quantities are listed in the bottom right
        of each panel. The sizes of \Lya near-zones for individual
        lines-of-sight are shown with the open diamonds, with the
        average for each parameter set shown by the filled diamonds.
        The solid curves correspond to $R_{\alpha}^{\rm max}$ for
        $\Delta_{\rm lim}=0.4$, the dashed curves correspond to
        $R_{\rm ion}$ for $\Delta = 1$ and the dotted line indicates
        the length of the simulated sight-line.  The filled grey
        circles and open circles in the central panel show the
        observed sizes of \Lya near-zones, as determined by  Fan et al.
        (2006b) and Wyithe et al. (2005) respectively.}
      \label{fig:sizes} \end{center} \end{minipage}
\end{figure*}

Finally, the bottom panel in figure~\ref{fig:los} shows the resulting
\Lya spectrum for a pixel size of $\sim 0.25\rm~\AA$.  A
signal-to-noise per pixel of $20$  at the continuum level has been adopted, with a
  constant read-out signal-to-noise of $100$ per pixel.  Spectra of
similar quality are already obtainable with current instruments
(\citealt{Becker05,Becker06}).  However, for the synthetic spectra
used in this paper we adopt the moderate spectral resolution shown in
figure~\ref{fig:example}, which is more typical of most of the
observational data.  The solid vertical lines in both panels in
figure~\ref{fig:example} mark the extent of the quasar \HII region at
$11.8$ Mpc.  Clearly the extent of the \Lya near-zone, $R_{\alpha}
\simeq 8.3$ Mpc, is some way behind the \HII I-front.  Note also that
the extent of the \Lyb and \Lya near-zones are the same even though
the volume weighted neutral fraction of the IGM in this line-of-sight
is $\sim 10^{-2}$ prior to the quasar switching on.  In this instance,
naively using equation~(\ref{eq:Rbeta}), one might expect the \Lyb
near-zone to be somewhat larger than the \Lya near-zone.  The reason
for the similarity in their size is clear on examining the density
distribution along this particular sight line, shown in the top panel
in figure~\ref{fig:los}.  The transmission peaks in the \Lya and \Lyb
spectra at $\sim 8.3$ Mpc correspond to the very low density void at
the same position in the density distribution.  Beyond this there are
several high density clumps which attenuate the quasar radiation field
and prevent any further transmission in either spectra. Consequently,
variations in the density distribution of the IGM, as well as the
potential insensitivity of the \Lya near-zone size to the neutral
hydrogen fraction, will make attempts to determine $f_{\rm HI}$ using
near-zone sizes difficult.  Combining the \Lya and \Lyb near-zone
sizes obviously helps, but this will not necessarily give the correct
extent of the quasar \HII region for any individual spectrum either.
As we will discuss in more detail later, reasonably large samples of
high quality data will be required to better determine the ionization
state of the surrounding IGM using near-zone sizes.  We further note
that the presence of the lower redshift \Lya forest in the \Lyb
spectrum clearly reduces the amplitude of the \Lyb transition and in
some cases will block transmission peaks from the \Lyb near-zone.

\subsection{The measured sizes of \Lya near-zones in cosmological density fields}

In figure~\ref{fig:sizes} we explore the dependence of \Lya near-zone
sizes on the ionization state of the surrounding IGM and the
luminosity and lifetime of quasars using realistic density
distributions drawn from our cosmological hydrodynamical simulation.
The left panel shows the dependence of \Lya near-zone sizes on the
volume weighted IGM neutral hydrogen fraction.  For each neutral
hydrogen fraction we have performed a full radiative transfer
calculation for the five different lines-of-sight drawn from our
simulation box at $z=6$.  The near-zone sizes were measured out to our
fiducial flux limit of $F_{\rm lim} = e^{-2.3} = 0.1$, and are shown
by the open diamonds in figure~\ref{fig:sizes}.  Note that some of
these data points fall on top of each other, giving the appearance of
less than five data points for some of the parameters.  The resolution
of each synthetic spectrum was degraded and noise was added to mimic
the observational data. The larger filled diamonds show the average
value of $R_{\alpha}$ for each parameter set.  The neutral hydrogen
fraction of the surrounding IGM was varied using UV background models
$0-10$, described in table~\ref{tab:preion}.  The lifetime and
luminosity of the quasar were fixed at $t_{\rm Q}=10^{7}\rm~yrs$ and
$\dot N = 2\times 10^{57} \rm~s^{-1}$, respectively.  The solid curve
shows $R_{\alpha}^{\rm max}$ assuming $\Delta_{\rm lim}=0.4$.  The
dashed curve shows $R_{\rm ion}$ assuming $\Delta = 1$ and the
horizontal dotted line corresponds to the length of each simulated
line-of-sight.

The first point to note about figure~\ref{fig:sizes} is the
considerable spread in the inferred near-zone sizes from one
line-of-sight to the next when all other parameters are fixed.  This
highlights the importance of taking into account the effect of a
realistic density field containing voids and overdense clumps on the
observed size and appearance of the near-zone.  In particular, as in
figure~\ref{fig:example}, it is the most underdense voids within the
quasar \HII region which set the observed extent of the
spectroscopically identified \Lya near-zone.

We can now investigate how the actual near-zone sizes relate to our
analytical estimates for $R_{\rm ion}$ and $R_{\alpha}^{\rm max}$
given in section~\ref{sec:stromprox}. Firstly, we consider the
effect of varying the surrounding IGM neutral hydrogen fraction, as
shown in the left hand panel of figure~\ref{fig:sizes}.   For large
values of the neutral hydrogen fraction the observed \Lya near-zone
size can be approximated by $R_{\rm ion}$, albeit with an offset from
our chosen value of $\Delta=1$ due to gas clumping. The near-zone size
increases with decreasing neutral fraction as $f_{\rm HI}^{-1/3}$, but
note also the slight deviation of the data point for an entirely
neutral IGM from the expected scaling due to the presence of an
extended damping wing from the Gunn-Peterson trough.  As expected the
observed sizes saturate at $R_{\alpha}^{\rm max}$ where they do not
increase further with decreasing neutral fraction until $f_{\rm HI}
\leq 10^{-3.5}$.  However, even for $f_{\rm HI}>0.1$ it is difficult
to determine the neutral fraction accurately from a small sample of
absorption spectra due to the rather large scatter in near-zone sizes.
Evaluating equation~(\ref{eq:timescale}) using $R_{\alpha}^{\rm max} =
7.85$, corresponding to $\Delta_{\rm lim}=0.4$ in
equation~(\ref{eq:Ralpha}), and $f_{\rm HI} \geq 0.1$ results in a
timescale for the \HII I-front to reach $R_{\alpha}$ greater than or
similar to the assumed quasar age of $t_{\rm Q}=10^{7} \rm~yrs$ if
$\Delta \geq 1$. For the chosen values of the quasar ionizing
luminosity and lifetime, saturation thus occurs at $f_{\rm HI} < 0.1$.
For larger quasar ionizing luminosity or longer lifetime the
saturation would occur at larger neutral fraction and {\it vice
  versa}.

For neutral hydrogen fractions smaller than $10^{-3.5}$, the \Lya
near-zone sizes measured using our definition (see section 2.2)
increase rapidly as transmission from the IGM ionized by the
metagalactic UV background rather than the quasar becomes more common.
Strictly speaking, these sizes no longer correspond to the regions
impacted by the quasar ionizing radiation.  With further decreasing
neutral fraction the near zone sizes rapidly approach the box-size of
the numerical simulation shown by the dotted line.  This is one of the
reasons why \cite{Fan06t} take the necessary step of smoothing their
\Lya spectra to obtain the \Lya near-zone sizes; not doing so makes
distinguishing between the many thin transmission peaks from the
highly ionized, underdense IGM and the true edge of the near-zone very
difficult.  However, smoothing will have the effect of attributing
transmission peaks from the IGM to the size of the quasar near-zone,
influencing the perceived evolution in smoothed near-zone sizes in the
regime where the IGM makes the transition from being optically thick
to optically thin.  One must be aware of this effect when interpreting
data which has been smoothed.  We discuss this in further detail in
section~\ref{sec:evolution}.

In the middle and right panels of figure~\ref{fig:sizes} we show the
dependence of the \Lya near-zone size on quasar ionizing luminosity
and quasar lifetime for an IGM neutral hydrogen fraction $f_{\rm HI}
\simeq 10^{-3}$ (UV background model $6$).  The dashed and solid
curves again show $R_{\rm ion}$ and $R_{\alpha}^{\rm max}$,
respectively. The simulated near-zone sizes scale with ionizing
luminosity as $R_{\alpha}\propto \dot N^{1/2}$ and are independent of
the lifetime of the quasar as expected for this neutral hydrogen
fraction, where $R_{\alpha} = R_{\alpha}^{\rm max}$.  Note that for a
significantly neutral IGM with $f_{\rm HI} \geq 0.1$, one would see an
evolution in the near-zone size roughly proportional to $t_{\rm
  Q}^{1/3}$ until $t_{\rm Q} \geq 10^{7} \rm~yrs$.  Overall, the
analytic approximations $R_{\rm ion}$ and $R_{\alpha}^{\rm max}$
appear to provide a reasonable guide for understanding the nature of
quasar near-zones for $10^{-3} \leq f_{\rm HI} \leq 1$.

The filled grey circles and open circles in the middle panel show the
observed sizes of \Lya near-zones as determined by \cite{Fan06t} and
\cite{Wyithe05} respectively.  There is good agreement between the
smaller near-zone sizes and the model, suggesting consistency with
$f_{\rm HI}\sim 10^{-3}$ in these cases.  The larger near-zone sizes,
predominantly at $z<6$, lie somewhat above the values from the
simulated spectra, but the uncertainties in the source luminosity, the
measured near-zone size and the near-zone size distribution due to
cosmic variance should be kept in mind.  We will come back to this
point later.  Note also again that the simulated spectra discussed
here have not been smoothed.  As we will see in
section~\ref{sec:evolution}, if we smooth the spectra this discrepancy
becomes larger.

\subsection{Comparing \Lya and \Lyb near-zone sizes}

\begin{figure}
\begin{center}
 
  \includegraphics[width=0.45\textwidth]{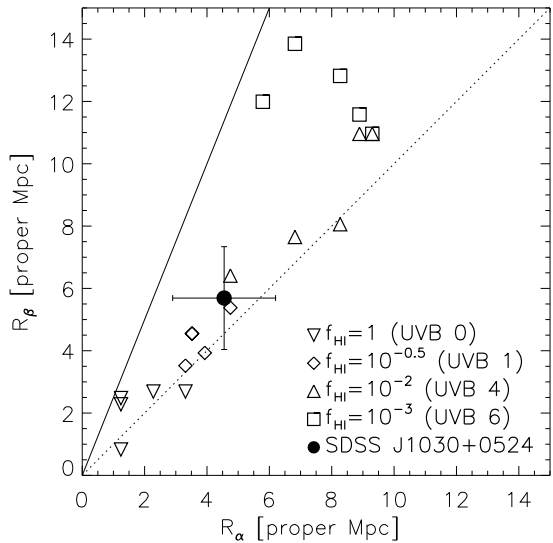}
  \caption{ The ratio of \Lya to \Lyb near-zone sizes for
  inhomogeneous IGM models with differing volume weighted neutral
  hydrogen fractions.  An ionizing photon production rate of $\dot N=2
  \times 10^{57}\rm~s^{-1}$ and age of $t_{\rm Q}=10^{7} \rm~yrs$
  has been assumed for the quasar.  The solid line shows $R_{\beta} =
  2.5R_{\alpha}$, which is the approximate upper limit to the size of
  the \Lyb near-zone predicted by equation~(\ref{eq:Rbeta}), while
  the dotted line shows $R_{\alpha}=R_{\beta}$.  The filled circle
  with error bars show the observationally determined sizes of the
  \Lya and \Lyb near-zones around $\rm J1030+0524$ (White et
  al. 2003).}  \label{fig:ratios}
\end{center} 
\end{figure}

In the last section we demonstrated that $R_{\alpha}$ alone is not
very sensitive to the ionization state of the surrounding IGM for
typical quasar ionizing luminosities and lifetimes.  This is due to
the saturation of the near-zone size for a low neutral hydrogen
fractions at $R_{\alpha}^{\rm max}$ and the large scatter in near-zone
sizes due to density fluctuations.  The relative size of $R_{\alpha}$
and $R_{\beta}$ is an independent probe of the ionization state which
stays useful when $R_{\alpha}$ is saturated at $R_{\alpha}^{\rm max}$.
In figure~\ref{fig:ratios} we compare the \Lya and \Lyb near-zone
sizes predicted by $20$ of the radiative transfer simulations with
progressively smaller IGM neutral hydrogen fractions. The solid line
shows $R_{\beta} = 2.5 R_{\alpha}$, which is the approximate upper
limit to the expected size of the \Lyb near-zone predicted by
equation~(\ref{eq:Rbeta}).  The dotted line shows
$R_{\alpha}=R_{\beta}$, the expected relation when the IGM is
substantially neutral.   As noted earlier, these theoretical limits ignore
the effect of foreground absorption from the lower redshift \Lya
forest on observed \Lyb near-zone sizes.  However, we do include
this effect in the synthetic spectra analysed here.

For $f_{\rm HI}\ge 10^{-2}$ the simulated spectra appear to follow the
$R_{\beta} = R_{\alpha}$ relation.  For small neutral hydrogen
fractions ($f_{\rm HI} \sim 10^{-3}$) the data exhibit an upward
deviation approaching but not reaching the $R_{\beta}= 2.5 R_{\alpha}$
relation.  This suggests that the inhomogeneous IGM, and to a lesser
extent the additional absorption from the \Lya forest in the \Lyb
spectrum, combined with limited spectral resolution, reduces the \Lyb
near-zone sizes more strongly than the \Lya near-zone sizes.  Hence,
for $f_{\rm HI} \sim 10^{-2}$ (UV background model 4) the \Lya and
\Lyb near-zone sizes can still be rather similar.

Note that for some of the completely neutral IGM models (UV
background model 0), $R_{\beta}$ is slightly larger than $R_{\alpha}$
due to the \Lya Gunn-Peterson trough damping wing ({\it e.g.}
\citealt{MesingerHaiman04}).  In the absence of this effect one would
expect $R_{\alpha} \simeq R_{\beta}$, since $R_{\alpha}$ has still not
reached its maximum and thus both near-zones trace the extent of
the \HII I-front.  Note further that  the time
scale for the \HII I-front to reach $R_{\beta}$ is around $16$ times
longer than the timescale to reach $R_{\alpha}$.  Consequently, the
\Lyb near-zone keeps increasing in size for some time after the
\Lya near-zone has ceased to expand.  This can be seen by
comparing the results for  UV background models 4 and 6.

The filled circle with error bars  show the  observationally determined
size of the \Lya and \Lyb near-zones around $\rm J1030+0524$
taken from White et al. (2003). We have assumed an uncertainty in the 
quasar redshift of $\Delta z = 0.03$. Its location close to the 
$R_{\alpha} = R_{\beta}$ line appears to suggest that  $f_{\rm HI}\ge 10^{-2}$. 
However, the large scatter in near-zone size makes it very difficult to
come to definite conclusions for individual systems. 

\section{The observed redshift evolution of near-zone sizes} \label{sec:evolution}
\subsection{The simulations}

In a recent paper \cite{Fan06t} (hereinafter F06b) present evidence for
evolution in the size of \Lya near-zone sizes with redshift.  The
conclusion of F06b using a sample of $16$ quasars, was that the average
\Lya near-zone size increases by a factor of $2.8$ from $z=6.4$ to
$z=5.7$ after correcting for quasar luminosity differences.  Assuming
that $R_{\alpha} \propto (1+z)^{-1}f_{\rm HI}^{-1/3}$, F06b note this
corresponds to $f_{\rm HI}$ decreasing by a factor of $\sim 14$.
Assuming a value of $f_{\rm HI}\sim 9.3 \times 10^{-5}$ at $z=5.7$
based on their direct optical depth measurements suggests $f_{\rm HI}
\sim 1.3 \times 10^{-3}$ at $z=6.4$.  However, as we have shown
already, the dependence of the \Lya near-zone size on $f_{\rm HI}$ is
not so straightforward.  We shall attempt to model this evolution here
using three different scenarios for the IGM neutral hydrogen fraction evolution
within our radiative transfer implementation.

In all three scenarios, we shall consider a quasar embedded in an
inhomogeneous hydrogen and helium IGM.  We set the quasar to have the
fiducial values $\dot N = 2 \times 10^{57} \rm~s^{-1}$ and $t_{\rm
  Q}=10^{7} \rm~yrs$.  As before, the IGM density distributions used
for the radiative transfer runs are drawn in different directions
around the $5$ most massive haloes in the simulation volume at
$z=[6.25,6,5.75]$, giving a total of $15$ different lines-of-sight for
comparison to the F06b data.  In the first scenario we simulate the
propagation of radiation from the source into the IGM at
$z=[6.25,6,5.75]$ using UV background models 6, 9 and 10 respectively.
These UV background models reproduce volume weighted neutral hydrogen
fractions which are consistent with estimates from the direct
Gunn-Peterson optical depth measurements of F06b.  This scenario
corresponds to a moderate evolution of the IGM neutral fraction with
redshift.  In the second scenario we use UV background models 1, 6 and
10 at $z=[6.25,6,5.75]$, respectively.  This scenario corresponds to a
rapid evolution in the IGM neutral hydrogen fraction above $z=6$,
where UV background model 1 corresponds to a substantially neutral IGM
at $z=6.25$, consistent with the lower limits on $f_{\rm HI}$ from
both \cite{MesingerHaiman04} and \cite{Wyithe05}.  In the third
scenario, we also adopt a moderate evolution in the neutral hydrogen
fraction, but in addition take into account the effect of
self-shielded dense clumps.  These systems are expected to be
responsible for Lyman limit systems, but our hydrodynamical
simulations do not fully resolve them.  To include them we use a
simple model for self-shielded clumps based on the model of
reionization in the post-overlap phase by \cite{MiraldaEscude00}. The
details of this model are described in appendix D.

\subsection{The relative sizes of \Lya near-zones and their evolution with redshift}

\begin{figure*}
  \centering 
  \begin{minipage}{180mm} 
    \begin{center}
           
      \psfig{figure=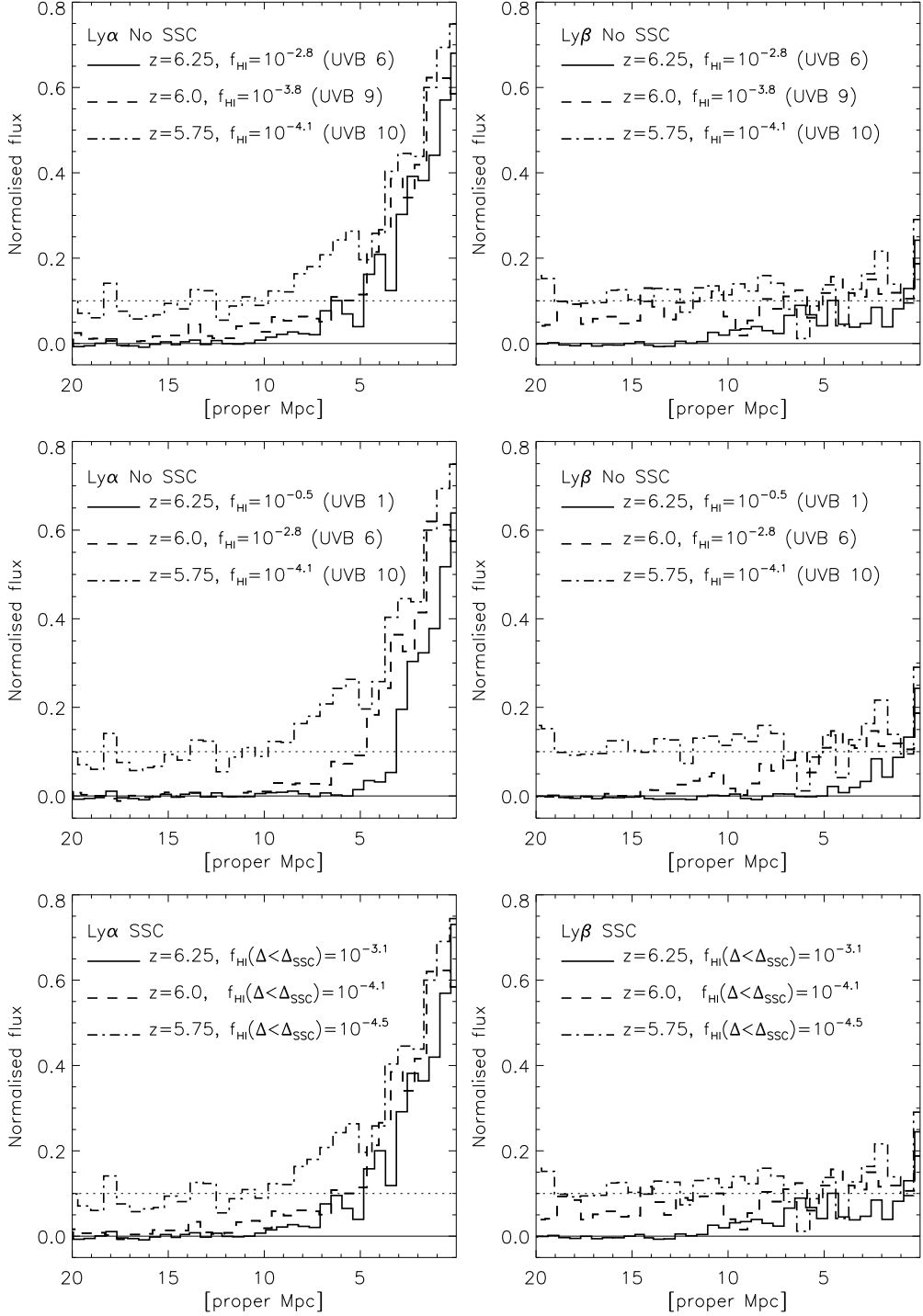,width=0.8\textwidth}
      \caption{ The evolution of \Lya and \Lyb near-zone size
        with redshift for different IGM neutral hydrogen fractions.
        Each panel shows the composite absorption profile constructed
        from $5$ lines-of-sight in redshift bins at $z=[6.25,6,5.75]$.
        The spectra have been degraded to resemble observational
        spectra and then smoothed to a resolution of $20\rm~\AA$.  The
        \Lya profiles are shown on the left hand side, with the \Lyb
        profiles on the right.  In all cases the quasar has an age of
        $t_{\rm Q}=10^{7}\rm~yrs$ and an ionizing photon production
        rate of $\dot N = 2 \times 10^{57}\rm~s^{-1}$.  {\it Top:}
        Near-zone sizes for an IGM ionized by a uniform
        metagalactic UV background, corresponding to the moderate
        evolution scenario. {\it Centre:} The UV background models are
        now chosen to produce a rapid evolution in the neutral
        hydrogen fraction.  {\it Bottom:} Near-zone sizes for an
        IGM which is highly ionized at low densities by a uniform UV
        background but is completely neutral above a fixed density
        threshold.  This mimics the behaviour of dense self-shielded
        clumps responsible for Lyman limit systems.  }
      \label{fig:fanplot}
     \end{center}
   \end{minipage}
\end{figure*}

As discussed in previous sections and by F06b, determining the absolute
sizes of the near-zones is plagued with a range of uncertainties. One
should, however, hope that these uncertainties depend weakly on
redshift.  Therefore, in this section we will firstly investigate the
redshift evolution in the relative sizes of \Lya near-zones found by
F06b.  We undertake a comparison of the absolute near-zone sizes with
the observations of F06b in the next section. The evolution of the
synthetic \Lya and \Lyb near-zone sizes with redshift for our three
scenarios is shown in figure~\ref{fig:fanplot}, similar to figure 13
in F06b.  We note however that figure 13 in F06b is not corrected for
the quasar luminosity, whereas for the simulations used in
figure~\ref{fig:fanplot} we have assumed a constant value of $\dot N =
2 \times 10^{57}\rm~s^{-1}$.  In addition, the spectra in each bin in
figure 13 of F06b are from a range of redshifts, rather than all at
the same redshift.

Following F06b, the simulated spectra are smoothed to a resolution of
$20\rm~\AA$ and the composite absorption profile of the $5$
lines-of-sight at each redshift is plotted.  To compare these data to
the observational results of F06b, we adopt their definition of the
\Lya near-zone.  This is defined as the extent of the smoothed
spectrum where the normalised flux is greater than $0.1$.  This
hopefully provides a more robust way of defining $R_{\alpha}$ when we
are considering examples where transmission from the thinning \Lya
forest affect the extent of the near-zones.

The data in the top panels in figure~\ref{fig:fanplot} correspond to
the moderate evolution scenario.  Between $z=5.75$ and $z=6.25$, the
\Lya near-zone size decreases by a factor of $2.1$.  This is slightly
less than the amount of evolution seen by F06b from $z=5.7$ to
$z=6.4$, but is consistent with the their data over the smaller
redshift range we consider.  Using the theoretical scaling $R_{\alpha}
\propto (1+z)^{-1}f_{\rm HI}^{-1/3}$, this would imply that the IGM
neutral hydrogen fraction has changed by a factor $\sim 7$ over the
range $5.75 \leq z \leq 6.25$.  However, the average volume weighted
IGM neutral hydrogen fraction used in each redshift bin in the top
panels is $f_{\rm HI} = [1.6 \times 10^{-3},1.5 \times 10^{-4},8.1
\times10^{-5}]$ at $z=[6.25,6,5.75]$.  Thus, the IGM neutral fraction
we use actually changes by a factor of $\sim 20$ over this range.
Note that there is little evolution in the composite \Lya near-zone
sizes between $z=6.25$ and $z=6$ because the sizes of the near-zones
are just leaving the saturated regime.  A substantial part of the
evolution of the near-zone size therefore occurred for the rather
small change in neutral fraction between $z=5.75$ and $z=6$, where the
IGM neutral hydrogen fraction changes by less than a factor of two. In
this redshift range the \Lya near-zones sizes are no longer near the
saturated regime, which occurs approximately for $ 10^{-3} \leq f_{\rm
  HI}\leq 10^{-1}$.  In this instance, transmission from the IGM
ionized by the UV background contributes to the \Lya near-zone size in
the smoothed spectra; the flux from the quasar at the edge of the
near-zone is comparable to the flux of the metagalactic UV background
ionizing the surrounding IGM.  Therefore, relatively small changes in
the highly ionized IGM neutral hydrogen fraction can produce a rapid
evolution in the observed sizes of \Lya near-zones using the F06b
definition of near-zone sizes.  This occurs at a neutral hydrogen
fraction of $f_{\rm HI} \sim 10^{-4}$.  Rapid evolution in the
smoothed \Lya near-zone sizes over the redshift range $5.75 \leq z
\leq 6.25$ is therefore not necessarily strong evidence for a
significantly neutral IGM just above $z=6$.  We briefly note that we
have tested this further by varying the flux limit used to define the
smoothed near-zone sizes.  For $F_{\rm lim}>0.1$, the redshift
evolution in the smoothed near-zone sizes is somewhat weaker. It
should be an interesting test to perform the same exercise on the
observed data.

In the top right panel we show the corresponding composite absorption
profiles for the \Lyb near-zones.  The level of transmission in the
\Lyb near-zone is $2-3$ times lower compared to the \Lya near-zone.
This is mainly due to foreground \Lya absorption from gas at lower
redshift.  The F06b definition for $R_{\alpha}$ is therefore not
appropriate as a definition for $R_{\beta}$. The transmission profile
of the \Lyb near-zone is much shallower, more extended and fluctuates
strongly.  The \Lyb near-zones all extend beyond the extent of the
corresponding \Lya near-zones, albeit by only around one Mpc at
$z=6.25$. This is expected since the optical depth for \Lyb absorption
is smaller and larger neutral hydrogen fractions can be probed.  

In the central panels we show the results for the rapid evolution
scenario, with $f_{\rm HI} = [3.1 \times 10^{-1},1.4 \times
10^{-3},8.1 \times10^{-5}]$ at $z=[6.25,6,5.75]$.  In this case, the
composite \Lya near-zone size decreases by factor of $3.1$ from
$z=5.75$ to $6.25$.  This is somewhat larger than the F06b observed
evolution.  Again the scaling $R_{\alpha} \propto (1+z)^{-1}f_{\rm
  HI}^{-1/3}$ would imply that $f_{\rm HI}$ has changed by a factor of
$24$, compared to an actual change by a factor of over $3800$.
However,  most of the near-zone size evolution in the
centre-left panel has again occurred for the rather small change in neutral
fraction between $z=5.75$ and $z=6$.  In the centre-right panel, at
$z=6$ the \Lyb transmission extends beyond the \Lya transmission as before.
In the $z=6.25$ case the \Lyb transmission extends
the same distance as the \Lya transmission, since in this case the
edge of the near-zones correspond to the quasar \HII I-front.  As
discussed earlier, the \Lya and \Lyb near-zones are expected to have a
similar size for an IGM with $f_{\rm HI} \geq 10^{-2}$ within our
models.  A comparison of the  \Lya and \Lyb near-zones sizes for a 
large sample of high quality  quasar spectra should thus tell 
whether the IGM is significantly neutral just above $z=6$ or not.

Finally, in the bottom panels we show the mean absorption profiles for
models which also take into account the effect of self-shielded clumps
(SSCs), as described in appendix D. These models should be a more
realistic representation of the post-overlap phase of reionization
(\citealt{MiraldaEscude00,Gnedin00}).  The average volume weighted
neutral hydrogen fraction in each redshift bin for these models is
$f_{\rm HI} = [7.4 \times 10^{-2},1.1 \times 10^{-2},4.3 \times
10^{-3}]$ at $z=[6.25,6,5.75]$.  Note, however, that the volume
weighted neutral fractions are considerably lower if the SSCs are
excluded, {\it i.e.} for the IGM with $\Delta < \Delta_{\rm SSC}$.
Excluding regions with $\Delta \geq \Delta_{\rm SSC}$ the neutral
hydrogen fractions in the highly ionized IGM are $f_{\rm
  HI}(\Delta<\Delta_{\rm SSC}) = [7.9\times 10^{-4},7.3 \times
10^{-5},3.5 \times 10^{-5}]$.  These values are similar to those used
for the moderate evolution scenario in the top panels.  Compared to
the models with no SSCs in the top panels, there is little difference
in the \Lya and \Lyb near-zones shown in the bottom panels.  Note that
although the volume weighted neutral hydrogen fraction is high in
these sight-lines, especially at $z=6.25$, the low density regions
which dominate the transmission features are still highly ionized.
Therefore, the inclusion of SSCs appears to have little effect on the
propagation of the quasar I-front. This is perhaps not too surprising
as the SSCs would have to contain a significant fraction of the total
hydrogen mass in neutral form in order induce a significant evolution
in the near-zone sizes with redshift. For a highly ionized IGM the
SSCs are therefore not important. This may, however, be different for
smaller neutral fractions and/or at higher redshift.

In summary, the relative sizes of \Lya near-zones and their evolution
with redshift are consistent with a moderate change in the IGM neutral
fraction.  Using an analysis similar to that performed on observed
near-zone spectra, we find even small changes in the neutral fraction
of a highly ionized IGM can produce a rapid evolution in the sizes of
\Lya near-zones if the flux from the quasar at the edge of the
near-zone is similar to the flux of the metagalactic UV background at
the Lyman limit.  However, because of the insensitivity of near-zones
to the neutral fraction and the scatter in near-zone sizes due to
fluctuations in the IGM density field, a substantially neutral IGM at
$z>6$ is also consistent with the data.  A detailed analysis of the
corresponding sizes of \Lyb near-zones should help to distinguish
between these possibilities.

\subsection{The absolute sizes of \Lya near-zones and their evolution with redshift}

\begin{figure}
\begin{center}
 
  \includegraphics[width=0.45\textwidth]{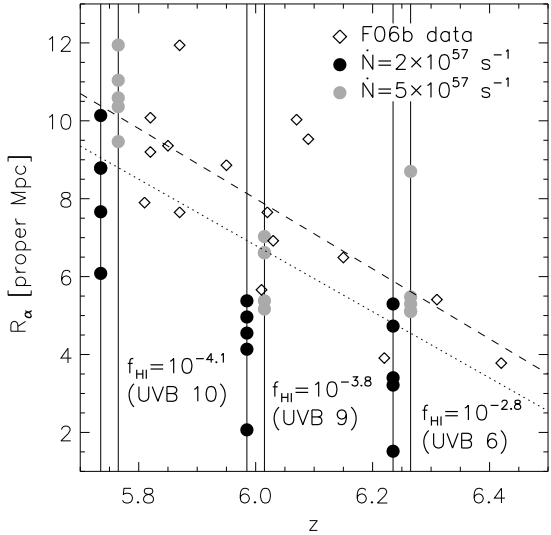}
  \caption{
    The sizes of \Lya near-zones as a function of redshift.  The
    filled black circles correspond to the individual \Lya near-zone
    sizes taken from the synthetic spectra in the moderate evolution
    scenario.  Here the quasar is assumed to have an ionizing
    luminosity of $\dot N = 2 \times 10^{57} \rm~s^{-1}$.  The
    observed correlation between \Lya near-zone size and redshift,
    determined empirically by Fan et al. (2006b), is shown with the
    dashed line.  This fit has been made with observed \Lya near-zone
    sizes rescaled to a common quasar ionizing luminosity of $\sim 1.9
    \times 10^{57} \rm~s^{-1}$, which are shown by the open diamonds.
    The simulation data lie slightly below this fit, although the
    evolutionary trend is roughly similar.  The filled grey circles
    correspond to the same synthetic sight-lines, except now the
    quasar ionizing luminosity is increased to $\dot N = 5 \times
    10^{57} \rm~s^{-1}$.  These data are in somewhat better agreement
    with the fit to the observational data. Note that the solid
    vertical lines are to guide the eye only.  In addition, only four
    data points are visible for the $\dot N=2\times 10^{57}\rm~s^{-1}$
    data at $z=6.25$ and $z=5.75$.  In these instances two near-zones
    have been measured to have the same size.}
    \label{fig:line}
\end{center} 
\end{figure} 

We now compare the absolute sizes of the synthetic \Lya near-zones
with the observations of F06b. In figure~\ref{fig:line} we plot the
individual \Lya near-zone sizes used in the moderate evolution
scenario against redshift as filled black circles, similar to figure
12 in F06b.  The dashed line is the empirically determined correlation
between $R_{\alpha}$ and redshift quoted by F06b. The dotted line is
the same relation assuming the observed near-zone redshifts were
overestimated by $\Delta z=0.02$, which is the uncertainty quoted by
F06b.  The observed \Lya near-zone sizes from which this fit is
derived in F06b, shown by the open diamonds in figure~\ref{fig:line},
are rescaled to a common absolute magnitude of $M_{1450}=-27$,
assuming $\dot N \propto M_{1450}$ and $R_{\alpha} \propto \dot
N^{1/3}$.  This is an uncertain correction and a rescaling by $\dot
N^{1/2}$ may actually be more appropriate.  Using the spectral energy
distribution given by equation~(\ref{eq:SED}), $M_{1450}=-27$
corresponds to $\dot N= 1.9 \times 10^{57} \rm~s^{-1}$, which is close
to the fiducial value of $\dot N = 2\times 10^{57}\rm ~s^{-1}$ adopted
for this study.  We also note briefly the linear fit to the
observational data shown in figure 12 of F06b is given by $R_{\alpha}
= 8 - 9(z-6)\rm~Mpc$, which is reproduced in our
figure~\ref{fig:line}.  This is slightly different to equation (24) in
F06b, which they quoted erroneously (X. Fan, private communication).

Interestingly, we obtain \Lya near-zone sizes which are 20-50 per cent
smaller than the observational data. The relative evolution and
scatter in the near-zone sizes is broadly consistent with the F06b
results. This may indicate that the  actual density and temperature 
distribution  of the IGM is somewhat different from that in our
numerical simulations and that the actual mean neutral fraction is
somewhat smaller than we have assumed for the simulated spectra.  
Another  obvious way of reconciling the synthetic near-zone sizes
of our moderate evolution scenario with the observational data is to
increase the ionizing flux in the the near-zone.  The filled grey
circles in figure~\ref{fig:line} correspond to the same synthetic
sight-lines, except now the quasar ionizing luminosity is set to $\dot
N = 5 \times 10^{57} \rm~s^{-1}$, a factor of 2.5 larger than before.
The near-zone sizes are now in better agreement with the observational
data.  However, it appears unlikely that the observed quasar
luminosities are underestimated by a factor of two to three. If an
enhanced ionizing luminosity really were required to reproduce the
absolute sizes of observed \Lya near-zones, this may instead be
attributed to star forming galaxies which are likely to cluster in the
dense environment surrounding the quasar.

Recently, \cite{Yu05a} suggested that the majority of the neutral
hydrogen surrounding a quasar may have already been ionized by stellar
sources before the quasar switches on. However, a factor three larger
ionizing luminosity appears again too large, considering quasars are
already extremely luminous objects.  There is, however, an important
effect which our simulations do not take into account which can
increase the combined ionizing flux from galaxies (and faint quasars)
in the near-zone without an enhanced ionizing luminosity from
individual galaxies.  The enhanced ionizing flux from the quasar in
the near-zone will ionize dense neutral regions which otherwise block
the ionizing radiation from other sources of ionizing photons.  This
leads to a significant increase of the ionizing photon mean free path
within the near-zone compared to the mean free path in the surrounding
IGM.  This will lead to an increase in the metagalactic UV background
flux due to the galaxies and possibly other, fainter quasars within
the near-zone. An accurate estimate of this effect is difficult but a
rough estimate can be obtained using our SSC model
(\citealt{MiraldaEscude00,FurlanettoOh05}, and also see appendix D).
An increase by a factor of two or three in the ionizing flux at the
edge of the near-zone appears plausible.

There is a range of other effects which may affect the estimated
near-zone sizes. However, they should be less important than the
effects discussed above. Spatial fluctuations in the UV background
expected around the end of reionization may influence the size of
observed near-zones. In addition, gas in highly overdense regions is
likely to be shocked, producing hot, highly ionized gas.  This has not
been taken account in our simulations. However, the volume filling
factor of this hot, ionized gas phase is expected to be small at
$z=6$. The difficulty of the data reduction may also contribute to the
discrepancy.  The continuum fitting of these highly absorbed spectra
in the region of the broad \Lya emission line is notoriously
difficult.  Unidentified emission lines could easily lead to a quasar
continuum which is placed too low at the relevant distance from the
systemic redshift.  This would result in an overestimate of the
transmission and could therefore result in an overestimate of the
observed near-zone sizes.

\section{Conclusions and discussion} \label{sec:concs}

We have presented a detailed investigation of the properties of
near-zones in the Lyman series absorption spectra of $z\simeq 6$
quasars, using analytical estimates and an accurate radiative transfer
scheme for a hydrogen and helium IGM combined with density
distributions drawn from a large hydrodynamical simulation of a \LCDM
universe.  From the radiative transfer simulations we have produced
realistic mock spectra which we have analysed in a similar manner to
the current observational data. The main conclusions of this study are
as follows.  \newline

\noindent
$\bullet$ For large IGM neutral hydrogen fractions the size of the
near-zone increases as $f_{\rm HI} ^{-1/3}$, as expected for an \HII
I-front expanding into a significantly neutral medium.  For typical
luminosities of observed $z\simeq 6$ quasars and plausible lifetimes,
the near-zone sizes saturate at a neutral fraction $f_{\rm HI} \approx
0.1$. For smaller neutral fractions the spectroscopically identified
near-zones are more akin to classical proximity zones within a highly
ionized IGM, and their size is largely independent of the neutral hydrogen
fraction of the surrounding IGM.  \newline

\noindent
$\bullet$ The density fluctuations in a cosmologically representative
matter distribution lead to considerable scatter in the size of the
near-zones, even for fixed assumptions about the ionization state of
the surrounding IGM. It is thus difficult to determine the IGM neutral
hydrogen fraction from a small number of observed spectra. \newline

\noindent
$\bullet$ Observed sizes of spectroscopically identified near-zones in
the \Lya region alone are a rather poor indicator of the ionization
state of the surrounding IGM.  Observed near-zone sizes can
furthermore substantially underestimate the size of the region
impacted by the ionizing radiation of the quasar.
\newline

\noindent
$\bullet$ For small neutral hydrogen fractions quasar \HeIII
I-fronts lag considerably behind their \HII counterparts if \HeII is
still abundant. This leads to a temperature jump behind the \HeIII
I-front which can be located  within the observed near-zone for a suitable
combination of IGM neutral fraction, quasar luminosity and quasar
lifetime.  Detailed study of the width of the absorption features in
the \Lya near-zone may therefore reveal the position of the \HeIII front.  This
could then be used together with the \Lya near-zone size to constrain the
neutral hydrogen fraction.\newline

\noindent
$\bullet$ The ratio of near-zone sizes in the \Lya and \Lyb region of
quasar spectra provides another independent probe of the ionization
state of the surrounding IGM.  For neutral fractions $f_{\rm HI} \ge
10^{-2}$ we find that the size of the \Lya and \Lyb near-zones are similar. For
smaller neutral fractions the size of the \Lyb region becomes
substantially larger, approaching the theoretical limit of $R_{\beta}
\simeq 2.5 R_{\alpha}$.  The extent of the \Lyb near-zone relative to
\Lya is blurred by the overlying \Lya forest at lower redshift.  It is
also very sensitive to the density distribution along the
line-of-sight.  For this reason the ratio of the sizes of the \Lya and
\Lyb near-zones approaches the saturated value at a neutral fraction
somewhat smaller ($f_{\rm HI} \la 10^{-2}$) than where
saturation sets in for the sizes of \Lya near-zones alone ($f_{\rm HI} \sim
10^{-1}$). The exact value of the threshold neutral hydrogen fractions
depends on the quasar luminosity and lifetime.\newline

\noindent
$\bullet$ The synthetic spectra produced from our radiative transfer
simulations exhibit \Lya near-zones which are $20-50$ per cent smaller
than the observed spectra for plausible assumptions for the ionizing
luminosity of the the quasar and the flux of the metagalactic UV
background in the surrounding IGM.  The discrepancy is resolved if the
ionizing flux at the edge of the near-zone is enhanced by a factor of
two to three. Such an enhancement is plausible given the expected
increase in the ionizing photon mean free path in the near-zone
compared to the mean free path in the surrounding IGM.  This would
produce a corresponding enhancement of the metagalactic UV background
within the near-zone from stellar emission and possibly fainter
quasars.  Alternatively, it may suggest that the IGM has a lower
neutral fraction than we assumed in our simulations.  \newline

\noindent
$\bullet$ The strong evolution of the observed sizes of \Lya
near-zones over the short redshift interval $5.7<z<6.4$ is consistent
with, but does not require, a rapid evolution of the neutral hydrogen
fraction of the surrounding IGM.  The observed evolution in the \Lya
near-zone size by around a factor of two is equally consistent with a
moderate evolution in the IGM neutral hydrogen fraction.  Relatively
small changes in the highly ionized IGM neutral hydrogen fraction can
produce a rapid evolution in the observed sizes of \Lya near-zones if
the ionizing flux from the quasar at the edge of the near-zone is
comparable to the flux of the metagalactic UV background at the Lyman
limit in the surrounding IGM.  In this case the rapid evolution can be
attributed to additional transmission from the thinning \Lya forest.
This proposition can be tested by varying the flux level used to
define the extent of the near-zone sizes. For larger flux levels the
evolution should decrease if the increasing transmission from the
general \Lya forest is responsible for part or most of the evolution.
\newline

\noindent
$\bullet$ The effect of self-shielded clumps on the observed sizes of
the \Lya near-zone is small, and the expected evolution of the incidence
rate of self-shielded clumps appears to contribute little to the
rapid evolution of near-zone sizes.  \newline

\noindent
We note here that we have not considered two further possibilities
which may influence the size of quasar near-zones: a fluctuating UV
background and additional sources of ionizing radiation, such as star
forming galaxies, clustered in the vicinity of the quasar host halo.
Both of these are likely to increase the observed near-zone sizes.
Additionally, observational constraints on near-zone sizes are still
somewhat uncertain, as are quasar ionizing luminosities and
lifetimes.  However, a large sample of high quality quasar
spectra with an accurate determination of the sizes of the \Lya
and \Lyb near-zones should improve the constraints on the
ionization state of the IGM at $z>6$ considerably.  If most quasars
exhibit \Lyb near-zones which are similar in size to the counterpart
\Lya near-zone, this should favour an IGM with $f_{\rm HI} \geq
10^{-2}$.  If \Lyb near-zones were instead generally 
larger than their \Lya counterparts this would provide evidence for 
an IGM with $f_{\rm HI} \leq 10^{-2}$.  The information locked within the
widths of absorption features in the \Lya near-zones may provide an
interesting additional probe of the ionization state of the IGM at
high redshift.

\section*{Acknowledgements}

We thank the referee, Xiaohui Fan, for helpful comments which
substantially improved the clarity of this paper and for kindly
providing us with the list of near-zone sizes and quasar luminosities
used in figures 1, 5 and 8.  This research was conducted in
cooperation with SGI/Intel utilising the Altix 3700 supercomputer
COSMOS at the Department of Applied Mathematics and Theoretical
Physics in Cambridge.  COSMOS is a UK-CCC facility which is supported
by HEFCE and PPARC.  This research was supported in part by PPARC and
the National Science Foundation under Grant No. PHY99-07949.

\label{lastpage}

\appendix

\section{Time delay effects on observed sizes of cosmological \HII regions}

The existence of a finite speed for the transmission of real
information alters the {\it observed} sizes and shapes of cosmological
\HII regions.  These time delay effects, although having their basis
in special relativity with the postulate that the speed of light in
empty space always has the same value, are purely geometrical rather
than relativistic phenomena which manifest themselves when the
I-fronts travel at relativistic speeds.  The renewed interest in
cosmological \HII regions as a probe of the epoch of reionization has
meant these effects have been well documented within the recent
literature, to which we refer the interested reader ({\it e.g.}
\citealt{White03,Yu05a,Yu05b,Shapiro05}) However, we briefly review
these results here, as they provide an important simplification to our
radiative transfer scheme.

One can easily show the expression for the radius, $R$, of a \HII
region expanding into a static, pure hydrogen medium of constant
density and temperature is ({\it e.g.} \citealt{Wyithe04,Shapiro05})

\begin{equation} \frac{dR}{dt} = c \left[ \frac{ \dot N - \frac{4}{3}
      \pi R^{3} \alpha_{\rm HII} n_{\rm H}^{2}}{ \dot N + 4\pi R^{2}
      f_{\rm HI} n_{\rm H} c - \frac{4}{3} \pi R^{3} \alpha_{\rm HII}
      n_{\rm H}^{2}  } \right], \label{eq:rel-Ifront} \end{equation}

\noindent
where $\dot N$ is the number of ionizing photons emitted by the
monochromatic source per unit time, $n_{\rm H}$ is the number density
of hydrogen atoms, $f_{\rm HI} = n_{\rm HI}/n_{\rm H}$ is the neutral
hydrogen fraction and $\alpha_{\rm HII}$ is the recombination
coefficient for ionized hydrogen.  This expression explicitly
takes into account the time delay for a photon to travel from the
source to the edge of the \HII region, thus setting the limiting
expansion speed of the \HII region to the speed of light.

Now consider the expansion of a \HII region outwards from the
ionizing source along the line of sight to an observer.  Any photons
instantaneously detected by the observer were emitted at the same {\it
retarded} time, $t_{\rm R} = t - R/c$, where $t$ is the time when the
photons reach a distance $R$ from the source.  Evaluating
equation~\ref{eq:rel-Ifront} at the retarded time yields

\begin{equation} \frac{dR}{dt_{\rm R}} = \frac{ \dot N - \frac{4}{3} \pi R^{3} 
    \alpha_{\rm HII} n_{\rm H}^{2}}{4 \pi R^{2} f_{\rm HI} n_{\rm H}},
    \label{eq:nonrel-Ifront} \end{equation}

\noindent
which is identical to equation~(\ref{eq:HIIfront}).  Thus, it is clear
for this particular geometrical configuration that the expansion speed
of the \HII region in the observer's frame is no longer limited to the
speed of light; it is as if one can completely ignore time delay
effects and assume real information is transmitted instantaneously.
This leads to the observation of an initial superluminal expansion
phase close to the ionizing source, when the number of ionizing
photons per hydrogen atom at the I-front greatly exceeds unity,
allowing it to propagate at relativistic speeds.  This result is
particularly useful in the context of our radiative transfer
simulations, which we discuss in the next section.

\section{Radiative transfer implementation}

Modelling the extent of cosmological \HII regions around high redshift
quasars, while consistently tracking local changes in the ionization
state and temperature of the gas due to the filtering of ionizing
radiation, requires a full treatment of radiative transfer through a
hydrogen and helium IGM.  In a cosmologically expanding medium
I-fronts from a luminous quasar are highly supersonic and thus
decoupled from the hydrodynamical response of the gas.  This allows
one to ignore the radiation hydrodynamics of this particular problem,
providing a significant simplification to our numerical scheme.  For
this purpose we use an updated version of the radiative transfer
algorithm of \cite{Bolton04}.  This algorithm is based on the explicit
photon conservation method of \cite{Abel99}, which has the desirable
property that photons are conserved almost independently of spatial
resolution, allowing the accurate determination of physically relevant
quantities with only moderate computational expense.  We briefly
outline the updated algorithm here and then present some tests of the
code.

\subsection{The cosmological model}

We adopt the following time-redshift relation for a flat Universe
dominated by a cosmological constant and non-relativistic matter
(\citealt{Peebles93})

\begin{equation} H_{0}t = \frac{2}{3(1-\Omega_{\rm m})^{1/2}} \sinh^{-1} 
\left[ \left( \frac{1-\Omega_{\rm m}}{\Omega_{\rm m}} \right)^{1/2}
  a^{3/2} \right ], \label{eq:redt} \end{equation}

\noindent
where $H_{0} = 100h~ \rm km~s^{-1}~Mpc^{-1}$ is the present day Hubble
parameter, $\Omega_{\rm m}$ is the matter density as a fraction of the
critical density and $a = (1+z)^{-1}$ is the cosmological scale
factor.  We use this relation to update redshift dependent quantities
at each successive timestep $\delta t$ within the radiative transfer
implementation.  Additionally, from equation~(\ref{eq:redt}) it is
straightforward to derive an expression for the Hubble parameter

\begin{equation} H(t) = \Omega_{\rm m}^{1/2}H_{0}\cosh 
  \left[ \frac{3H_{0}}{2} (1- \Omega_{\rm m})^{1/2}t \right] a^{-3/2}.
  \label{eq:hubble} \end{equation}

\subsection{The ionizing radiation field}

Consider a source which emits $\dot N$ ionizing photons per unit time,
such that

\[ \dot N = \int_{\nu_{\rm HI}}^{\infty} \dot N_{\nu} d\nu = \int_{\nu_{\rm HI}}^{\infty} \frac{L_{\nu}}{h_{\rm p}\nu} d\nu, \]

\noindent
where $L_{\nu}~d\nu$ is the luminosity of the source within the frequency
interval $\nu \rightarrow \nu+d\nu$, defined as

\[ L_{\nu}~d\nu = L_{\rm HI} \left( \frac{\nu}{\nu_{\rm HI}} \right)^{- \alpha_{\rm s}} d\nu. \]

\noindent
The source has a power law spectrum with index $\alpha_{\rm s}$,
typical of quasars, normalised by $L_{\rm HI}$ at the \HI ionization
threshold frequency, $\nu_{\rm HI}$.

The probabilities of an ionizing photon emitted by the source then
being absorbed by H~$\rm \scriptstyle I$, \HeI and He~$\rm
\scriptstyle II$, respectively, are (\citealt{Bolton04})

\[ P_{\rm HI} =  p_{\rm HI}q_{\rm HeI}q_{\rm HeII}(1- {\rm e}^{-\tau_{\nu}^{\rm tot}})/D, \]

\[ P_{\rm HeI} = q_{\rm HI}p_{\rm HeI}q_{\rm HeII}(1- {\rm e}^{-\tau_{\nu}^{\rm tot}})/D, \]

\[ P_{\rm HeII} = q_{\rm HI}q_{\rm HeI}p_{\rm HeII}(1- {\rm e}^{-\tau_{\nu}^{\rm tot}})/D, \]

\noindent where $p_{\rm i} = 1- {\rm e}^{-\tau_{\nu}^{\rm i}}$,
$q_{\rm i} = {\rm e}^{-\tau_{\nu}^{\rm i}}$, $\tau_{\nu}^{\rm tot} =
\tau_{\nu}^{\rm HI}+\tau_{\nu}^{\rm HeI}+ \tau_{\nu}^{\rm HeII}$,
$\tau_{\nu}^{i}$ is the optical depth for species $i$ and $D=p_{\rm
HI}q_{\rm HeI}q_{\rm HeII}+q_{\rm HI} p_{\rm HeI}q_{\rm HeII}+q_{\rm
HI}q_{\rm HeI}p_{\rm HeII}$.

If we now invoke a uniformly discretised space grid with elements of
size $\delta R$ along the line-of-sight to the source, at any given
timestep $\delta t$, the photo-ionization rate per unit volume $\rm
[s^{-1}~cm^{-3}]$ of species $i$ in grid element $l$ is

\[ n_{\rm i}^{\rm l}\Gamma_{\rm i}^{\rm l} =  \frac{1}{V^{\rm l}} 
\int_{\nu_{\rm i}}^{\infty} \dot N_{\nu}^{\rm l-1} P_{\rm i}~d\nu, \]

\noindent
where $V^{\rm l}$ is the volume of element $l$ and $\nu_{\rm i}$ is
the ionization threshold frequency for species $i$.  If we assume the
luminosity of the source is constant and allow an additional
contribution to the ionization rate from a diffuse UV background, the
above equation can be rewritten as

\begin{equation} n_{\rm i}^{\rm l}\Gamma_{\rm i}^{\rm l} = \left[  \frac{1}{4 \pi R^{2} \delta R } 
\sum_{k} \frac{L_{\nu_{\rm k}}}{h_{\rm p}\nu_{\rm k}} P_{\rm i}e^{-T_{\nu_{\rm
  k}}} \right] + n_{\rm i}^{\rm l}\Gamma_{\rm i}^{\rm b}, \end{equation}

\noindent
which is the equation we shall solve for the ionization rate per unit
volume for species $\rm i$.  Here we have optimally discretised the
frequency integration into $20$ logarithmically spaced intervals $k$
between $\nu_{\rm i}$ and $10\nu_{\rm i}$, $R$ is the distance of
element $l$ from the source, $\Gamma_{\rm i}^{\rm b}$ is the optically
thin photo-ionization rate due to the diffuse metagalactic UV
background and the transmission factor, $T_{\nu}$, is defined as

\[ T_{\nu} =  
\cases{ 0 &($l=0$),\cr \noalign{\vskip3pt} \sum_{j=0}^{l-l}
  (\sigma_{\rm HI}n_{\rm HI}^{j} + \sigma_{\rm HeI}n_{\rm HeI}^{j} +
  \sigma_{\rm HeII}n_{\rm HeII}^{j})\delta R &($l>0$).\cr} \]

\noindent
Similarly, the photo-heating rate per unit volume $\rm
[erg~s^{-1}~cm^{-3}]$ for species $i$ is given by

\begin{equation}  n_{\rm i}^{\rm l}{\epsilon}_{\rm i}^{\rm l} = \left[ \frac{1}{4 \pi R^{2} \delta R}
\sum_{\rm k} \frac{L_{\nu_{\rm k}}}{h_{\rm p}\nu_{\rm k}} P_{\rm i} h_{\rm p}(\nu_{\rm
k} - \nu_{\rm i}) e^{-T_{\nu_{\rm k}}} \right] + n_{\rm i}^{\rm
l}\epsilon_{\rm i}^{\rm b}, \end{equation}

\noindent
such that the total photo-heating rate per unit volume in element $l$ is given by
${\mathscr H}_{\rm tot}^{l} = n_{\rm HI}\epsilon_{\rm HI}^{\rm l} +
n_{\rm HeI}\epsilon_{\rm HeI}^{\rm l} + n_{\rm HeII}\epsilon_{\rm
  HeII}^{\rm l}$ and $\epsilon_{\rm i}^{\rm b}$ is the optically
thin photo-heating rate due to the diffuse UV background.

In general, for each timestep $\delta t$ the number densities $n_{\rm
  i}$ in the expression for the transmission factor $T_{\nu}$ are
evaluated in each grid cell $j$ at the same {\it retarded} time
$t_{\rm R} = t - j \delta R/c$, where $t$ is the time it takes a light
signal to reach cell $j$.  In principle, an I-front propagating
outwards from the source is then limited to travel no faster than the
speed of light since $n_{\rm i}$ are only evaluated when $t \geq
j\delta R/c$.  However, in practice spatial discretisation means this
is not always the case; an I-front is assumed to propagate
instantaneously across a cell width $\delta R$.  Even for relatively
high spatial resolution, this enables the expansion rate of the
ionized region to exceed $c$.  This means that to accurately compute
the position of an I-front one must either increase the spatial
resolution to a prohibitively high level or artificially impose the
limit $R \leq ct$.  At first glance, the latter option would seem to
be preferable.  However, this has the undesirable effect of violating
photon conservation, so that although one correctly predicts the
position of the I-front ({\it e.g.} \citealt{Abel99}), one must again
resort to substantially higher spatial resolution (such that each cell
in the space grid is optically thin) to compute physically relevant
quantities within the region where $R \leq ct$ is imposed.  This is a
problem for simulating quasar near-zones in Lyman series absorption,
since the absorption spectrum is sensitive to the exact values of the neutral
fraction and temperature of the gas near the source.  Fortunately, we
have already shown that in the frame of an observer the expansion rate of
the I-front follows equation~(\ref{eq:nonrel-Ifront}).  Since we are
interested in the sizes of {\it observed} \HII regions, we may
evaluate the expression for $T_{\nu}$ assuming the photons are
transmitted instantaneously, removing the need to impose the condition
$R\leq ct$ while at the same time explicitly conserving photons.  This
enables us to determine the ionized fraction and temperature of the
gas close to the ionizing source accurately for relatively coarse
spatial resolution, substantially reducing the computational expense
of the simulations.

\subsection{Computing the ionized fraction and temperature}

At every timestep $\delta t$, in each spatial grid element $\delta
R$, the ionization state and temperature of atomic hydrogen and helium
must be computed by solving a set of four coupled first order ordinary
differential equations.  The first three determine the abundances of
the three ionized species

\begin{equation} \frac{dn_{\rm HII}}{dt} = n_{\rm HI}(\Gamma_{\rm HI} 
+ n_{\rm e}\Gamma_{\rm eHI}) - n_{\rm HII}n_{\rm e}\alpha_{\rm HII},
 \label{eqn:H1} \end{equation}

\[ \frac{dn_{\rm HeII}}{dt} = n_{\rm HeI}(\Gamma_{\rm HeI} + n_{e}\Gamma_{\rm eHeI})
+ n_{\rm HeIII}n_{\rm e}\alpha_{\rm HeIII} \]
\begin{equation} \hspace{10mm} - n_{\rm HeII}(\Gamma_{\rm HeII} 
+ n_{\rm e}\Gamma_{\rm eHeII} + n_{\rm e}\alpha_{\rm HeII}),
\label{eqn:He1} \end{equation}

\begin{equation} \frac{dn_{\rm HeIII}}{dt} = n_{\rm HeII}(\Gamma_{\rm HeII} 
+ n_{\rm e}\Gamma_{\rm eHeII}) - n_{\rm HeIII}n_{\rm e}\alpha_{\rm
  HeIII}, \label{eqn:He2} \end{equation}

\noindent
with the closing conditions

\[ n_{\rm HI} = n_{\rm H} - n_{\rm HII}, \]
\[ n_{\rm HeI} = \frac{Y}{4(1-Y)}n_{\rm H} - n_{\rm HeII} - n_{\rm HeIII}, \]
\[ n_{\rm e} = n_{\rm HII} + n_{\rm HeII} + 2n_{\rm HeIII}.\]

\noindent
Here $n_{\rm i}$ denotes the number density $[\rm cm^{-3}]$ of species
$i$, $\Gamma_{\rm i}$ $[\rm s^{-1}]$ is the photo-ionization rate,
$\Gamma_{\rm ei}$ $[\rm cm^{3}~s^{-1}]$ is the collisional ionization
rate, $\alpha_{\rm i}$ $[\rm cm^{3}~s^{-1}]$ is the total radiative
recombination coefficient and $Y$ is the helium mass fraction, which we
take to be $Y=0.24$ ({\it e.g.} \citealt{OliveSkillman04}).

Additionally, the temperature must also be computed by solving a
fourth differential equation

\begin{equation} \frac{dT}{dt} = \frac{(\gamma-1) \mu m_{\rm H}}{k_{\rm B} \rho}
  [{\mathscr H}_{\rm tot}  - \Lambda(n_{\rm i},T)] - 2H(t)T,
  \label{eqn:T} \end{equation}

\noindent
where $\mu$ is the mean molecular weight of the gas, $\gamma = 5/3$,
$\Lambda(n_{\rm i},T)$ $\rm [erg~s^{-1}~cm^{-3}]$ is the radiative
cooling function, $H(t)$ is the Hubble parameter given by
equation~(\ref{eq:hubble}) and all other symbols have their usual
meanings.  The last term on the right hand side of equation~(\ref{eqn:T})
is the contribution to cooling by the adiabatic expansion of the
Universe.

We solve equations~(\ref{eqn:H1})-(\ref{eqn:T}) using a first order
implicit integration scheme which provides a good balance between
accuracy and speed (\citealt{Anninos97}).  We find that to solve these
equations accurately, one must typically adopt a timestep comparable
to the hydrogen ionization timescale, $t_{\rm ion} = 1/\Gamma_{\rm HI}$, which
can become very small in close proximity to a luminous source.  On
occasion this time interval is smaller than the main timestep we use
to track the position of the I-front (see section~\ref{sec:restest}
for details).  When this is the case we sub-cycle over each main time
step using smaller time intervals based on the ionization timescale
to solve for the ionic abundances.  Once the rate of change in the
electron number density is less than $10^{-8}$, we switch to solving
the ionic abundances assuming the gas is in ionization equilibrium
({\it e.g.} \citealt{Katz96}).  This leaves only the differential
equation for the temperature to be integrated, allowing much larger
timesteps to be taken which greatly speeds up the calculation.

\subsection{Cross-sections and rates}

The cross-sections and rates adopted in our radiative transfer scheme
are listed below.  Note that we use case A instead of case B
recombination coefficients for this study.  For optically thick
hydrogen and helium, one can approximately take into account the
moderate boost to the photo-ionization rates due to diffuse radiation
emitted by recombining ions using case B recombination coefficients.
This is the so-called on-the-spot approximation (\citealt{Spitzer78}).
However, any gas seen in \Lya absorption in our simulated spectra
typically lies behind the source I-fronts and is thus optically thin.
In addition, for typical gas densities the recombination timescale is
considerably longer than the canonical quasar lifetime.  In these
circumstances we find case A recombination coefficients are the
appropriate choice.

\vspace{3mm}

\noindent
(1) {\it Photo-ionization cross-sections} $[\rm cm^{2}]$ (\citealt{Osterbrock89}):

\[ \sigma_{\rm HI} = 6.30 \times 10^{-18}[ 1.34(\nu/\nu_{\rm HI})^{-2.99} 
-0.34(\nu/\nu_{\rm HI})^{-3.99}], \]

\[ \sigma_{\rm HeI} = 7.03 \times 10^{-18} [ 1.66(\nu/\nu_{\rm HeI})^{-2.05} 
-0.66(\nu/\nu_{\rm HeI})^{-3.05}], \]

\[ \sigma_{\rm HeII} = 1.50 \times 10^{-18} [ 1.34 (\nu/\nu_{\rm HeII})^{-2.99} 
-0.34(\nu/\nu_{\rm HeII})^{-3.99}]. \]

\noindent
(2) {\it Optically thin photo-ionization rates} $[\rm
s^{-1}]$:

\[ \Gamma_{\rm HI}^{\rm b} = 1.27 \times 10^{-11} J_{-21}^{\rm b}(\alpha_{\rm b} + 3)^{-1}, \]

\[ \Gamma_{\rm HeI}^{\rm b} = 1.51 \times 10^{-11} (0.553)^{\alpha_{\rm b}} J_{-21}^{\rm b}(\alpha_{\rm b} + 2)^{-1}, \]

\[ \Gamma_{\rm HeII}^{\rm b} = 3.03 \times 10^{-12} (0.250)^{\alpha_{\rm b}} J_{-21}^{\rm b}(\alpha_{\rm b} + 3)^{-1}, \]

\noindent
where the specific intensity of the metagalactic radiation field $\rm
[erg~s^{-1}~cm^{-2}sr^{-1}Hz^{-1}]$ is defined as

\[ J_{\nu}^{\rm b} = J_{-21}^{\rm b} \times 10^{-21} \left( \frac{\nu}{\nu_{\rm HI}} \right)^{-\alpha_{\rm b}}. \]

\noindent
The radiation field is assumed to have a power law spectrum with index
$\alpha_{\rm b}$, normalised by $J_{-21}^{\rm b}$ at the $\rm HI$
ionization threshold.

\vspace{3mm}

\noindent
(3) {\it Optically thin photo-heating rates} $[\rm
erg~s^{-1}]$:

\[ \epsilon_{\rm HI}^{\rm b} = 2.94 \times 10^{-22} J_{-21}^{\rm b}[(\alpha_{\rm b}+2)(\alpha_{\rm b}+3)]^{-1}, \]

\[ \epsilon_{\rm HeI}^{\rm b} = 6.48 \times 10^{-22} (0.553)^{\alpha_{\rm b}} J_{-21}^{\rm b}
[(\alpha_{\rm b}+1)(\alpha_{\rm b}+2)]^{-1}, \]

\[ \epsilon_{\rm HeII}^{\rm b} = 2.80 \times 10^{-22} (0.250)^{\alpha_{\rm b}} J_{-21}^{\rm b} 
[(\alpha_{\rm b}+2)(\alpha_{\rm b}+3)]^{-1}. \]

\noindent
(2) {\it Recombination rates} $[\rm cm^{3}~s^{-1}]$ (\citealt{Abel97}):
\noindent
\[ \alpha_{\rm HII} = \exp(-28.6130338 - 0.72411256\ln(\tilde{T}) \]
\vspace{-3mm}
\[ \hspace{7mm} - 2.02604473 \times 10^{-2}\ln(\tilde{T})^{2} - 2.38086188 \times 10^{-3}\ln(\tilde{T})^{3} \] 
\vspace{-3mm}
\[ \hspace{7mm} - 3.21260521 \times 10^{-4}\ln(\tilde{T})^{4} - 1.42150291 \times 10^{-5}\ln(\tilde{T})^{5} \]
\vspace{-3mm}
\[ \hspace{7mm} + 4.98910892 \times 10^{-6}\ln(\tilde{T})^{6} + 5.75561414 \times 10^{-7}\ln(\tilde{T})^{7} \] 
\vspace{-3mm}
\[ \hspace{7mm} - 1.85676704 \times 10^{-8}\ln(\tilde{T})^{8} - 3.07113524 \times 10^{-9}\ln(\tilde{T})^{9}, \]

\[ \alpha_{\rm HeII} = \alpha_{\rm HeII}^{\rm r} + \alpha_{\rm HeII}^{\rm d}, \]

\[ \alpha_{\rm HeII}^{\rm r} = 3.925 \times 10^{-13} \tilde{T}^{-0.6353}, \]

\[ \alpha_{\rm HeII}^{\rm d} = 1.544 \times 10^{-9} \tilde{T}^{-1.5}
{\rm e}^{-48.596/\tilde{T}}(0.3 + {\rm e}^{8.1/\tilde{T}}), \]

\[\alpha_{\rm HeIII} = 2 \alpha_{\rm HII}(\tilde{T}/4), \]

\noindent
where $\tilde{T}$ is the gas temperature in electron volts.
\vspace{3mm}

\noindent
(4) {\it Collisional ionization rates} $[\rm cm^{3}~ s^{-1}]$
(\citealt{Theuns98}):

\[ \Gamma_{\rm eHI} = 1.17\times 10^{-10} T^{0.5} {\rm e}^{-157809.1/T} [ 1 + T_{5}^{1/2}]^{-1},  \]

\[ \Gamma_{\rm eHeI} = 4.76\times 10^{-11} T^{0.5} {\rm e}^{-285335.4/T} [ 1 + T_{5}^{1/2}]^{-1}, \]

\[ \Gamma_{\rm eHeII} = 1.14\times 10^{-11} T^{0.5} {\rm e}^{-631515.0/T} [ 1 + T_{5}^{1/2}]^{-1}, \]

\noindent
where $T_{5} = T/ 10^{5} \rm~K$.  \vspace{3mm}

\noindent
The following radiative cooling processes contribute towards the
cooling function $\Lambda(n_{\rm i},T)$: \vspace{3mm}

\noindent
(5) {\it Recombination cooling} $[\rm erg~s^{-1}~cm^{-3}]$:

\[ \Lambda_{\rm HII}^{\rm rec} = 1.036\times 10^{-16} T \alpha_{\rm HII} n_{\rm e} n_{\rm HII},\]

\[ \Lambda_{\rm HeII}^{\rm rec} =  (1.036\times 10^{-16} T \alpha_{\rm HeII}^{\rm r} + 6.526 \times 10^{-11}\alpha_{\rm HeII}^{\rm d}) n_{\rm e} n_{\rm HeII},\]

\[ \Lambda_{\rm HeIII}^{\rm rec} =   1.036\times 10^{-16} T \alpha_{\rm HeIII}  n_{\rm e} n_{\rm HeIII}.\]

\noindent
(6) {\it Collisional ionization cooling} $[\rm erg~s^{-1}~cm^{-3}]$
(\citealt{Theuns98}):

\[ \Lambda_{\rm eHI} =  2.18 \times 10^{-11}\Gamma_{\rm eHI} n_{\rm e} n_{\rm HI},\]

\[ \Lambda_{\rm eHeI} =  3.94 \times 10^{-11} \Gamma_{\rm eHeI} n_{\rm e} n_{\rm HeI},\]

\[ \Lambda_{\rm eHeII} =   8.72 \times 10^{-11} \Gamma_{\rm eHeII} n_{\rm e} n_{\rm HeII}.\]

\noindent
(7) {\it Collisional excitation cooling} $[\rm erg~s^{-1}~cm^{-3}]$
(\citealt{Cen92}):

\[  \Lambda_{\rm HI}^{\rm ex} = 7.50\times 10^{-19} {\rm e}^{-118348/T} [ 1 + T_{5}^{1/2}]^{-1} 
n_{\rm e} n_{\rm HI}, \]

\[ \Lambda_{\rm HeI}^{\rm ex} = 9.10 \times 10^{-27} T^{-0.1687} {\rm e}^{-13179.0/T}[ 1 + T_{5}^{1/2}]^{-1} n_{\rm e}^{2} n_{\rm HeI},\] 

\[ \Lambda_{\rm HeII}^{\rm ex} = 5.54 \times 10^{-17} T^{-0.397} {\rm e}^{-473638/T}[ 1 + T_{5}^{1/2}]^{-1} n_{\rm e} n_{\rm HeII}.\] 

\noindent
(8) {\it Bremsstrahlung} $[\rm erg~s^{-1}~cm^{-3}]$
(\citealt{Cen92}):

\[ \Lambda_{\rm ff} = 1.43 \times 10^{-27} T^{1/2} g_{\rm ff} ( n_{\rm HII} + n_{\rm HeII} + 4 n_{\rm HeIII}) n_{\rm e}, \]

\noindent
where the Gaunt factor for free-free emission, $g_{\rm ff}$, is given
by

\[ g_{\rm ff} =  1.1 + 0.34 \exp[-(5.5 - \log_{10}(T)]^{2}. \]
\noindent
(9) {\it Inverse Compton cooling} $[\rm erg~s^{-1}~cm^{-3}]$
(\citealt{Peebles71}):

\[ \Lambda_{\rm c} = 5.65 \times 10^{-36} [T - 2.73(1+z)](1+z)^{4} n_{\rm e}.\]

\section{Code tests}

\begin{figure*}
  \centering \begin{minipage}{180mm} \begin{center}

      \psfig{figure=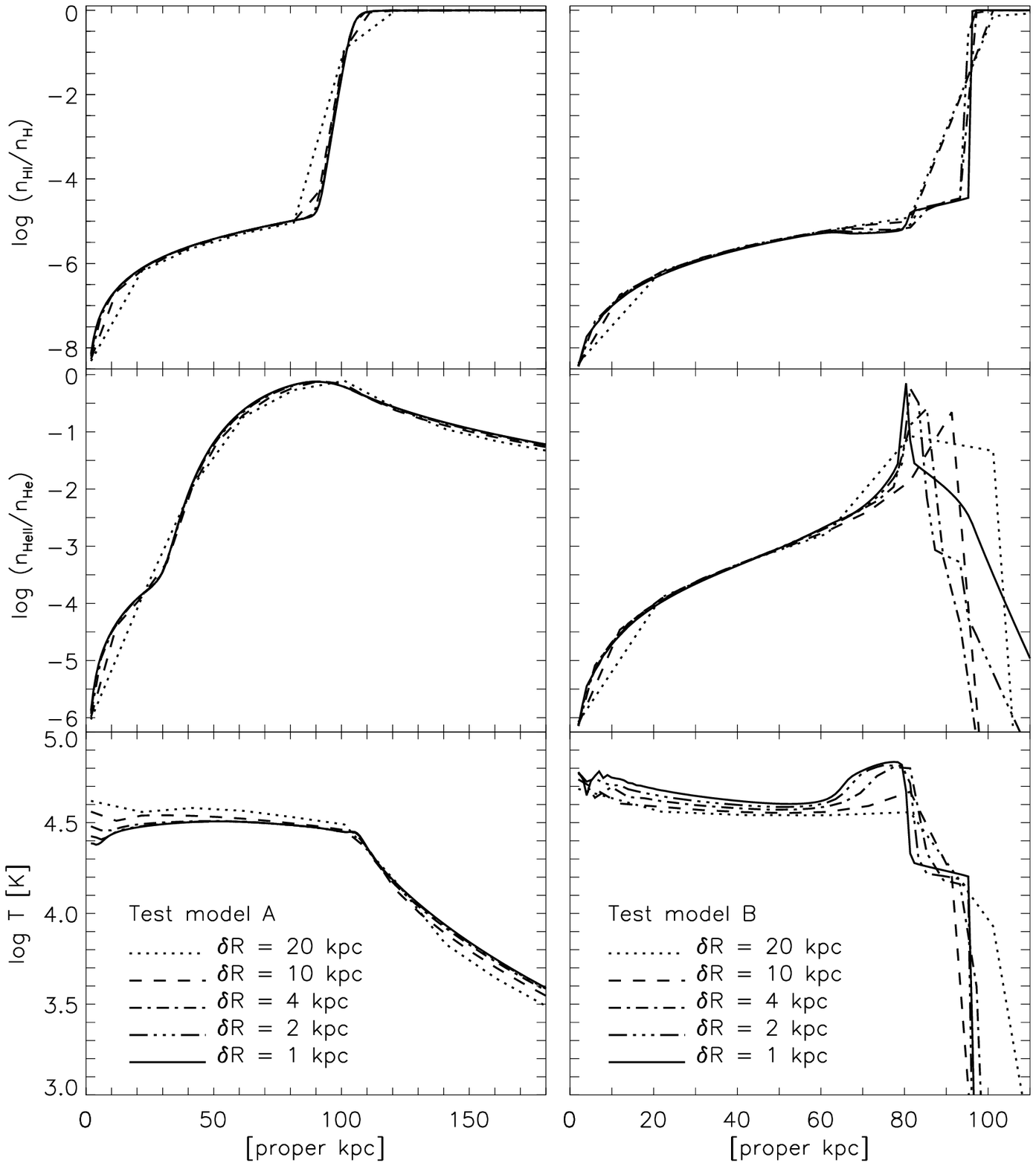,width=0.85\textwidth}
      \caption{ Test of spatial convergence for our radiative transfer
        scheme.  The left hand panels show the fractional abundances
        of H~$\rm \scriptstyle I$, \HeII and the gas temperature for
        model A using the cell sizes labelled on the plot.  The right
        hand panels show the same quantities for model B.}
      \label{fig:resolutionx}
      \end{center} \end{minipage}
\end{figure*}

\begin{figure*}
  \centering \begin{minipage}{180mm} \begin{center}

      \psfig{figure=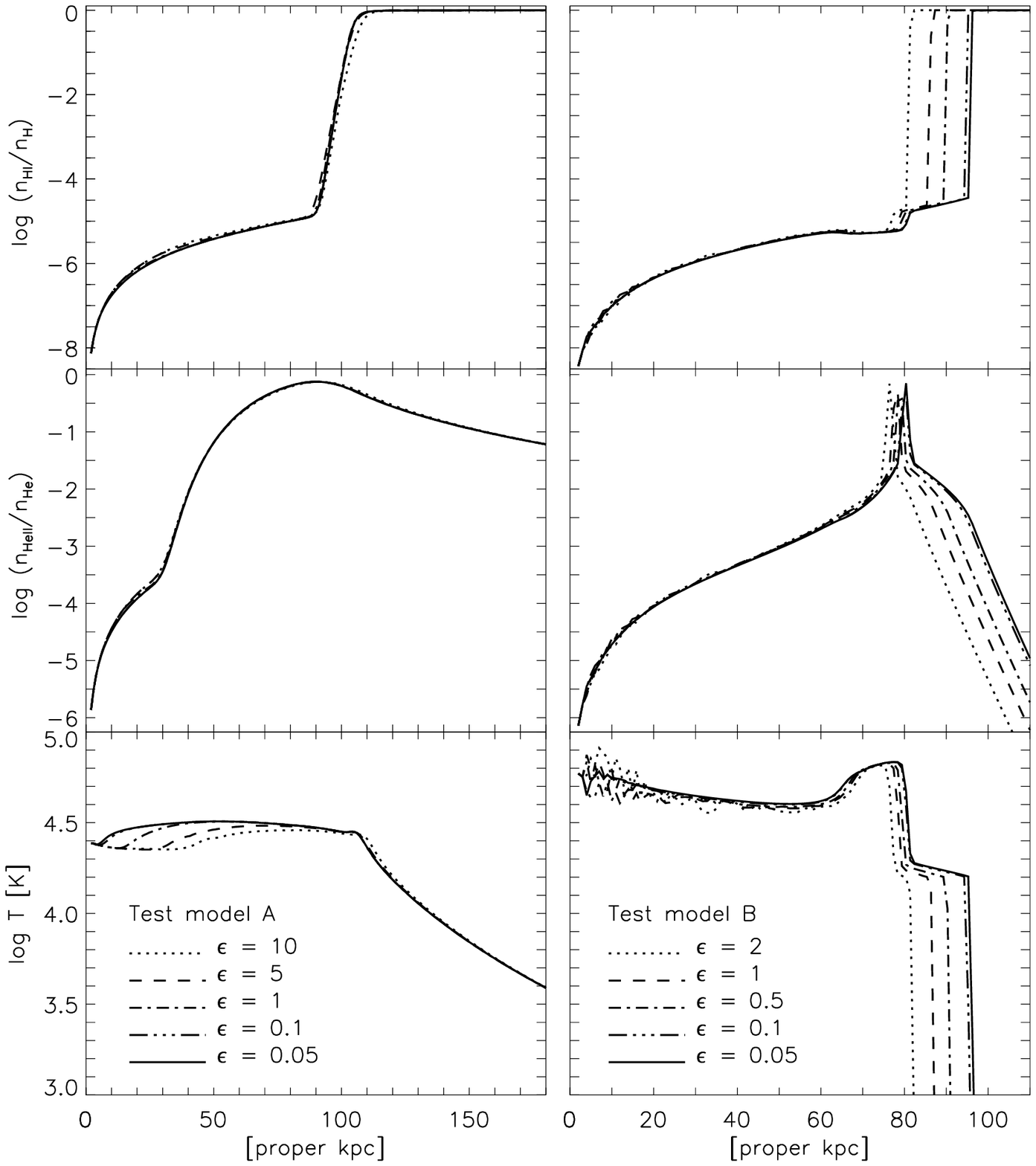,width=0.85\textwidth}
      \caption{  Test of temporal convergence for our radiative
        transfer scheme.  The left hand panels show the fractional
        abundances of H~$\rm \scriptstyle I$, \HeII and the gas
        temperature for model A using the Courant conditions labelled
        on the plot.  The right hand panels show the same quantities
        for model B. }
      \label{fig:resolutiont} \end{center} \end{minipage}
\end{figure*}

Exploring the limitations of any numerical model is essential for
providing well motivated results.  To this end we present some
selected tests of the radiative transfer implementation we use for our
simulations.  The test models are summarised in table~\ref{tab:models}

\begin{table} 
  \centering \caption{ Radiative transfer test models.  For each model
  the columns list the ionizing photon production rate of the
  source $\dot N$, the spectral index of the source $\alpha_{\rm s}$,
  the hydrogen number density $n_{\rm H}$, the mass fraction of helium
  $Y$ and the age of the source $t_{\rm s}$. Models C and D are for a
  monochromatic source, therefore no source spectral indices are
  listed.}

  \begin{tabular}{c|c|c|c|c|c} \hline Model & $\dot N \rm~[s^{-1}] $ &
$\alpha_{\rm s}$ & $n_{\rm H} \rm~[cm^{-3}]$ & $Y$ & $t_{\rm s}
\rm~[yrs] $\\   \hline A  &  $5 \times 10^{53}$ & $1.5$ & $7 \times
10^{-5}$ & 0.24 & $ 10^{6}$  \\ B  & $5 \times 10^{56}$ &
$1.5$ & $7 \times 10^{-2}$ & 0.24 & $ 10^{6}$  \\ C  & $1
\times 10^{55}$  & $-$ & $7 \times 10^{-5}$ & 0 & $ 10^{7}$
\\ D  & $1 \times 10^{59}$ & $-$  & $7 \times 10^{-2}$ & 0 &  $10^{7}$  \\ \hline
\label{tab:models}
\end{tabular}
\end{table}

\subsection{Spatial and temporal resolution requirements} \label{sec:restest}

We firstly present some convergence tests for the spatial and temporal
resolution of our radiative transfer scheme.  Our algorithm has the
useful property that photons are conserved explicitly independent of
spatial resolution ({\it e.g.} \citealt{Abel99,Bolton04,Mellema06}).
Consequently, the ionized fraction and temperature are computed
accurately even when the cells on the space grid are very optically
thick, speeding the calculation up considerably.

The optical depth to neutral hydrogen in each cell of width $\delta R$
on the space grid is given by $\delta \tau_{\rm HI} = \sigma_{\rm HI}
n_{\rm HI} \delta R$.  Employing fiducial values, this can be written
as

\begin{equation} \delta \tau_{\rm HI} =  13.6 \left( \frac{n_{\rm HI}}{7 \times 10^{-5} \rm~ cm^{-3}}
\right)\left ( \frac{\delta R}{ 10 \rm~ kpc} \right ). \end{equation}

\noindent
Note that in the case of a hydrogen and helium IGM, the optical depth
to \HeI and \HeII per cell is less than that for \HI due to the lower
number density of these species and, in the case of He~$\rm \scriptstyle II$, a smaller
photo-ionization cross section.  In this case the \HI optical depth
per cell is the correct proxy to use for the spatial resolution.

The results of a spatial resolution test using models A and B are
shown in figure~\ref{fig:resolutionx}.  We have adopted the same time
step size,$\epsilon=0.05$, in both of these models.  We shall discuss
this in more detail shortly. In both models, the propagation of
ionizing radiation into a hydrogen and helium medium of uniform
density is computed for a source with a power-law spectrum situated $2
\rm~kpc$ from the edge of the simulation grid.  The left panels show
the fractional abundances of H~$\rm \scriptstyle I$, \HeII and the gas temperature for
model A, with a hydrogen number density of $7\times 10^{-5}~\rm
cm^{-3}$, comparable to the mean density of hydrogen at $z=6$.  Even
adopting cell sizes in which the hydrogen is initially very optically
thick we find all quantities are recovered with a good degree of
convergence. The right hand panels show the same results for model B,
which has a hydrogen number density $1000$ times greater than model A.
Again, the degree of convergence for such optically thick cells is
quite remarkable.  Note there is evidence that the gas temperature has
not fully converged at the lowest spatial resolutions, especially
close to the source and near the \HeIII ionization front where there
are rapid changes in the energy input per ionization.  However, we
find a spatial resolution of $\delta R \simeq 10 \rm~kpc$ provides a
good degree of accuracy for the range of parameters we shall consider.

The second factor we need to consider is the temporal resolution.  As
already discussed in the previous section, potentially the shortest
timescale we need to consider is the hydrogen ionization timescale
$t_{\rm ion}=1/\Gamma_{\rm HI}$.  However, this timescale varies locally in the
simulation.  More generally the timestep size we are interested in is
the cell crossing time for real information, which we parameterise
using the Courant condition $\epsilon$, defined as

\begin{equation} \epsilon = \frac{c \delta t}{\delta R}. \end{equation}

\noindent
We adopt this as the main timestep size of our radiative transfer
scheme.  Note here that the time resolution is dependent on the spatial
resolution within our implementation.  Again, using typical values
this can be rewritten as

\begin{equation} \delta t = 3261.6 \left( \frac{\epsilon}{0.1} \right) 
\left( \frac{\delta R}{10 \rm~ kpc} \right) \rm~ yrs. \end{equation}

\noindent
Figure~\ref{fig:resolutiont} shows the results of our temporal
resolution tests for models A and B.  We have adopted a cell size of
$\delta R = 1 \rm~kpc$ for all of these test runs.  It is immediately
clear from model A that a timestep equal to or shorter than the cell
crossing time ($\epsilon \leq 1$) is required to achieve numerical
convergence.  For model B which has a higher gas density, a stricter
requirement $\epsilon \leq 0.1$ is required to produce reasonably
converged results.  This is because the rate of change in the optical
depth of each cell is greater per timestep.  For fully
three-dimensional cosmological radiative transfer with multiple
sources this timestep size would be somewhat prohibitive and could be
improved upon by using a form of adaptive timestepping based on the
I-front velocity (\citealt{Shapiro04}) or time averaged optical depths
(\citealt{Mellema06}).  However, for our purpose this simple time
stepping scheme is perfectly adequate.  Consequently, we adopt a value
of $\epsilon = 0.1$ for our simulations.

\subsection{Pure hydrogen isothermal \HII region}

Modelling the expansion of a \HII region into neutral hydrogen gas of
uniform density ({\it e.g.} \citealt{Stromgren39}) is one of the few
radiative transfer problems which has an exact analytical solution
with which one can compare simulation output (\citealt{Iliev06}).  We
consider two test cases (C and D) which simulate the propagation of an
\HII I-front into an initially neutral pure hydrogen medium of uniform
density and temperature $(T=1.5 \times 10^{4} \rm~K)$.  The ionizing
source is assumed to be steady and monochromatic, emitting $\dot N$
photons per unit time at the Lyman limit.

\begin{figure}
\begin{center}
  
  \includegraphics[width=0.45\textwidth]{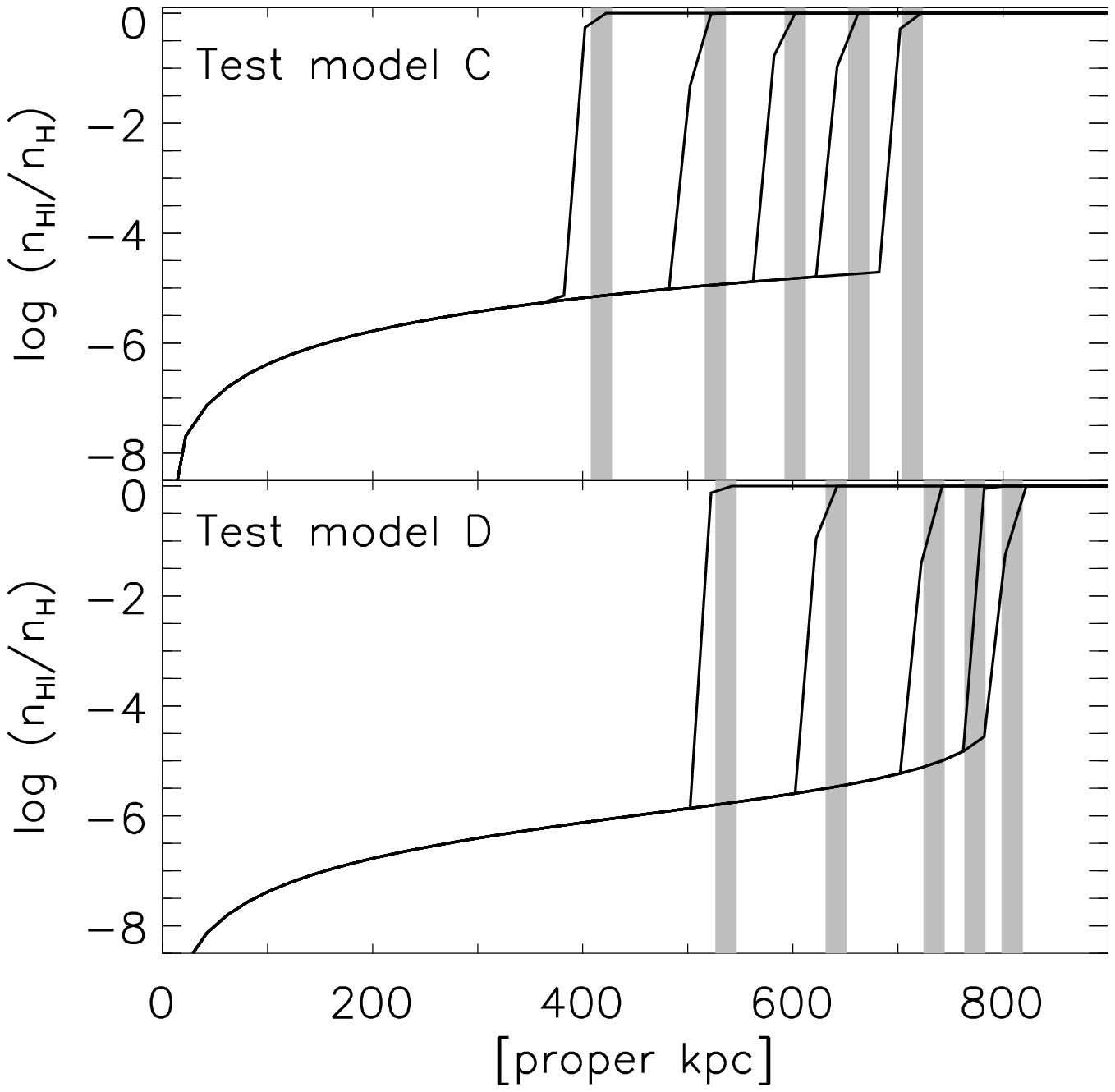}
  \caption{ Comparison of test models C (upper panel) and D (lower
  panel) to the analytical solution for the size of a \HII region in
  a pure hydrogen medium of constant temperature and density.  The
  solid lines show the numerical solution for the \HI fraction at
  $t/10^{7} \rm~yrs = [0.2,0.4,0.6,0.8,1.0]$ for model C and $t/10^{7}
  \rm~yrs = [0.05,0.1,0.2,0.3,1.0]$ for model D.  The centre of the solid
  grey strips give the analytical solution with the width of the
  strips representing the cell size of the simulation.}
  \label{fig:stromgren}
\end{center} 
\end{figure} 

We compare the results of test models C and D to the classical
analytical solution, equation~(\ref{eq:stromgren}), in
figure~\ref{fig:stromgren}.  The solid lines show the numerical
solution for the \HI fraction at $t/10^{7} \rm~yrs =
[0.2,0.4,0.6,0.8,1.0]$ for model C and $t/10^{7} \rm~yrs =
[0.05,0.1,0.2,0.3,1.0]$ for model D.  The solid grey strips give the
analytical solution with a width equal to the cell size of the
simulation, and hence the uncertainty in the numerical result.  For
this test we adopt a cell size of $\delta R = 20\rm~kpc$, with
$\epsilon=0.1$ for model C and $\epsilon=0.01$ for model D.  The
agreement between the numerical and analytical results is very good,
which suggests that we are calculating the extent of the
\HII regions correctly within our radiative transfer implementation.
Note that by $t/10^{7} \rm~yrs = 1.0$ model D has reached its
Str{\"o}mgren radius.

\section{A simple model for self-shielded regions} \label{sec:SSCs}

After the low density gas in the IGM has been reionized, dense,
self-shielded clumps (SSCs) are thought to set the mean free path for
ionizing photons ({\it e.g.}  \citealt{MiraldaEscude00}, hereinafter
MHR00, \citealt{MiraldaEscude03}).  We adopt a simple model to include
these SSCs within our simulations.  As noted by \cite{FurlanettoOh05},
one may approximate the density at which hydrogen in a dense clump
becomes self-shielded.  For a clump of normalised density
$\Delta=\rho_{\rm b}/\bar \rho_{\rm b}$ in ionization equilibrium with
the UV background, its column density is approximately
(\citealt{Schaye01}),

\begin{equation} N_{\rm HI} = 4.5 \times 10^{14} \frac{\Delta^{3/2}}{\Gamma_{-12}^{\rm b}} \left( \frac{T}{10^{4} \rm~K} \right)^{0.2} \left( \frac{1+z}{7} \right)^{9/2} \rm~cm^{-2}, \end{equation}

\noindent
where $\Gamma_{-12}^{\rm b} = \Gamma_{\rm HI}^{\rm
  b}/10^{-12}\rm~s^{-1}$ is the normalised metagalactic hydrogen
ionization rate and $T$ is the gas temperature.  The clump will become
optically thick to ionizing radiation once $N_{\rm HI} \geq
1/\sigma_{\rm HI} = 1.6 \times 10^{17} \rm~cm^{-2}$.  Adopting this
value for $N_{\rm HI}$ and rearranging yields the expected normalised
density threshold for a SSC immersed in the metagalactic radiation
field

\begin{equation} \Delta_{\rm SSC}  \simeq 49.5 \left( \frac{T}{10^{4}\rm~K} \right)^{0.13} \left( \frac{1+z}{7} \right)^{-3} \left(\Gamma_{-12}^{\rm b}\right)^{2/3}. \label{eq:clumps2} \end{equation}

\noindent
For UV background models $6$, $9$ and $10$, which produce IGM neutral
fractions consistent with the observational data of \cite{Fan06t},
this implies $\Delta_{\rm SSC} = [4.1,21.2,37.5]$ at
$z=[6.25,6,5.75]$.  Ideally, we would implement this self-shielding
for $\Delta \geq \Delta_{\rm SSC}$ within our radiative transfer
simulations in the same way as MHR00, by assuming all regions with
$\Delta \geq \Delta_{\rm SSC}$ remain neutral while still optically
thick to ionizing radiation.  However, although we have shown that
the resolution of our hydrodynamical simulation should be sufficient
for correctly modelling the propagation of ionizing radiation, it
still does not quite reach the required mass resolution to reproduce
the correct number of clumps with $\Delta \geq \Delta_{\rm SSC}$.
MHR00 overcome this problem by using an IGM density distribution drawn
from a hydrodynamical simulation, with the addition of a power-law
high density tail to account for the lack of high density clumps.  The
mean free path for ionizing photons in their model is given by

\begin{equation} \lambda_{\rm mfp} = \frac{60}{H(z)} \left[ 1- F_{\rm V}(\Delta<\Delta_{\rm SSC}) \right]^{-2/3} \rm~Mpc, \end{equation}

\noindent
where $H(z)$ is the Hubble parameter and $F_{\rm V}(\Delta <
\Delta_{\rm SSC})$ is the fraction of the IGM volume with $\Delta <
\Delta_{\rm SSC}$, computed from the volume weighted density
distribution corrected for high densities.
Assuming $\lambda_{\rm mfp}$ is similar to the mean distance between
SSCs, which is expected to be a reasonable approximation
(\citealt{FurlanettoOh05}), the number of SSCs per unit redshift with
$\Delta \geq [4.1,21.2,37.5]$ at $z=[6.25,6,5.75]$ using the MHR00 model is

\begin{equation} \frac{dN_{\rm SSC}}{dz} \simeq \frac{1}{\lambda_{\rm mfp}} \frac{c}{H(z)(1+z)} = [52.5,9.0,5.1]. \label{eq:clumps} \end{equation}

\noindent
In comparison, the number of clumps per unit redshift with $\Delta
\geq [4.1,21.2,37.5]$ at $z=[6.25,6,5.75]$ within our $400^{3}$
simulation volume is around a factor of two less than this.
Therefore, when setting all clumps with $\Delta \geq \Delta_{\rm SSC}$
to be neutral within our simulations using UVB models 6, 9 and 10, we
underpredict the mean free path for ionizing photons by about a factor
of two.  To compensate for this, we lower the density threshold at
which a clump becomes self-shielded, thus reproducing the required
number of SSCs per unit redshift given by the MHR00 model, as listed
in equation~(\ref{eq:clumps}).  The relevant density thresholds in our
$400^{3}$ simulation which reproduce the required number of SSCs per
unit redshift are $\Delta_{\rm SSC} = [2.7,10.0,20.0]$ at
$z=[6.25,6,5.75]$.  Thus, our simulations using $\Delta_{\rm SSC} =
[2.7,10.0,20.0]$ and UV background models 6, 9 and 10 at
$z=[6.25,6,5.75]$ should be equivalent to using $\Delta_{\rm SSC} =
[4.1,21.2,37.5]$ with $\lambda_{\rm mfp} = [1.09,6.99,12.80] \rm~Mpc$
and $\Gamma_{-12}^{\rm b} = [0.028,0.280,0.560]$.  Assuming ionization
equilibrium at $\Delta=1$ and $T=2 \times 10^{4}\rm~K$, these
ionization rates correspond to $f_{\rm HI}= [7.3\times 10^{-4},7.3
\times 10^{-5},3.6 \times 10^{-5}]$.  However, we note that
simulations incorporating radiative transfer which correctly resolve
the number density of SSCs are needed to fully address their effect on
observed near-zone sizes.

\end{document}